\documentclass[11pt,a4paper]{article}
\pdfoutput=1
\usepackage{jcappub}
\usepackage{amsmath,amssymb}
\usepackage{float}
\def\vector#1{\mbox{\boldmath $#1$}}
\def\vk{\mbox{\boldmath $k$}}
\def\vx{\mbox{\boldmath $x$}}

\def\vq{\mbox{\boldmath $q$}}

\def\gaij{\gamma_{ij}}

\def\biari{-\frac{H^3}{4\epsilon k_2}}

\def\biarii{-\frac{H^3}{4\epsilon k_1}}
\def\binashiq{\frac{H^2}{q^3}}

\def\pidot{\dot{\pi}}
\def\gtil{\tilde{g}}
\def\parti{\partial_{i}}
\def\partj{\partial_{j}}

\def\ttil{\tilde{t}}
\def\sigdot{\dot{\sigma}}
\def\zetadot{\dot{\zeta}}
\def\hankel{H_{\nu}^{(1)}}
\def\hankelc{H_{\nu}^{(1)*}}
\def\hankelm{H_{i\mu}^{(1)}}
\def\hankelmc{H_{i\mu}^{(1)*}}

\def\sigmatip{e^{-\pi\textrm{Im}(\nu)}\frac{\pi}{4}H^2}
\def\csch{\textrm{csch}}
\def\coth{\textrm{coth}}
\def\cosh{\textrm{cosh}}
\def\sinh{\textrm{sinh}}

\def\pfq{{}_2F_2}
\def\bessel{J_{i\mu}}
\def\besselc{J_{-i\mu}}
\def\cnp{c_{n}^{+}}
\def\cnm{c_{n}^{-}}
  \makeatletter
    
    \@addtoreset{equation}{section}
  \makeatother

\title{Heavy Particle Signatures in Cosmological Correlation Functions with Tensor Modes }

\author{Ryo Saito  and}
\author{Takahiro Kubota}

\affiliation{Department of Physics, Osaka University,  Toyonaka, Osaka 560-0043, Japan}
\emailAdd{saito@het.phys.sci.osaka-u.ac.jp}
\emailAdd{takahirokubota859@hotmail.com}
\abstract
{
We explore the possibility to make use of cosmological data to look for signatures of unknown heavy  particles 
whose masses are on the order of the Hubble parameter during the time of inflation.   To be more specific 
we take up the quasi-single field inflation model, in which the isocurvaton $\sigma $ is supposed to be 
the heavy particle. We study correlation functions involving both scalar ($\zeta $) and tensor ($\gamma $) 
perturbations and search for imprints of the $\sigma$-particle effects. We make use of the technique 
of the effective field theory for inflation to derive the $\zeta \sigma $ and $\gamma \zeta \sigma $ 
couplings. With these couplings we compute the effects due to $\sigma $ to the power spectrum  
$\langle \zeta \zeta \rangle $ and  correlations  
$\langle \gamma^{s} \zeta \zeta \rangle$ and 
$\langle \gamma^{s_{1}} \gamma ^{s_{2}} \zeta \zeta \rangle $, where $s$, $s_{1}$ and $s_{2}$ are the 
polarization indices of gravitons. Numerical analyses of the $\sigma$-mass effects to these correlations 
are presented.  It is argued that future precise observations of these correlations could make it possible 
to measure the $\sigma$-mass and the strength of the $\zeta \sigma$ and $\gamma \zeta \sigma$ couplings. 
As an extension to the $N$-graviton case we also compute the correlations 
$\langle \gamma ^{s_{1}} \cdots \gamma ^{s_{N}} \zeta \zeta \rangle $
and
$\langle \gamma ^{s_{1}} \cdots \cdots \gamma ^{s_{2N}} \zeta \zeta \rangle $
and their $\sigma$-mass effects. It is suggested that larger $N$ correlation functions are useful 
to probe larger $\sigma$-mass . 
}


\begin{document}
\maketitle

%%%%%%%%%%%%%%%%%%%%%%%%%%%%%%%%%%%%%%%%%%%%%
\section{Introduction}

It has been by now well accepted that the  inflationary expansion  in the early Universe is the key 
to solve various problems in the Big Bang cosmology. 
The methods to obtain the information during inflation have been developed in various ways. 
One of the  far-sighted theoretical bases  
 has been laid down   by Maldacena \cite{Mal}, who applied 
standard  quantum field theories to single-field inflation cosmology and computed the three-point functions 
involving  primordial 
fluctuations, $\zeta$ (scalar fluctuation) and  $\gaij$ (tensor fluctuation or `graviton'). 
A special attention was paid to  the three-point function of $\zeta$,  because it tells us 
much about the deviation from the Gaussian features of the cosmic state, often referred to as 
`non-Gaussianity' \cite{Komatsu}.

Another  intriguing and  closely related approach was put forward   by Cheung {\sl et al.} \cite{Effective}, 
who constructed the `Effective Field Theory (EFT) 
of Inflation' (see also Refs. \cite{Effective2} and \cite{Effective3}). 
In this theory, the scalar fluctuation $\zeta$ is interpreted as the 
 Nambu-Goldstone mode  associated with the spontaneous breaking of 
the time diffeomorphism invariance. 
A symmetry argument allows us to consider 
%more terms in a Lagrangian than Maldacena's, so we can discuss 
a more general framework  than Maldacena's by using a technique analogous to the low-energy pion physics.

Although general single field inflationary models have been explored to a considerable extent, 
there exists another avenue of investigation  in inflationary models that contain multiple fields
\cite{sasakitanaka, bernardeau, langlois, gong}. 
A class of  models is such that, in the space of multi-fields,   there is a flat slow-roll direction 
and all others are  isocurvature directions associated with isocurvatons whose masses are on  
the order of the Hubble parameter $H$.   This class of models is called `quasi-single field inflation'
and Chen and Wang \cite{Quasi} have pointed out that the massive isocurvaton, which they 
denoted by $\sigma $,  may have important observable effects on cosmological  perturbation. 
The magnitude of the Hubble parameter during inflation may be inferred by the formula
\begin{equation}
H \approx 2.5 \times 10^{13}\: \left ( \frac{r}{0.01}  \right )^{1/2} \quad {\rm GeV}\:, 
\end{equation}
where $r$ is the so-called tensor-to-scalar  ratio.
If quasi-single field inflation models were realized in Nature and if $r \sim 0.01$, then we 
would be  able to 
get  information of unknown particles whose masses are on the order of 
$10^{13} - 10^{14}\:\: {\rm GeV}$, which is far beyond the 
energy scale to be reached by terrestrial accelerators in the immediate future.

In their recent seminal paper, Noumi, Yamaguchi and Yokoyama \cite{Noumi} investigated the effects 
due to the heavy particle $\sigma$ applying the EFT method to the quasi-single field inflation. 
(See also Ref. \cite{tong}.)
They studied the three-point function $\langle \zeta \zeta \zeta \rangle $  and in particular its 
so-called squeezed limit, 
exploring the possibility of reading off the existence of the heavy particle $\sigma $.  
Imprints of new particles on the primordial cosmological fluctuations were also investigated 
in Ref. \cite{Cosmo}, emphasizing the role of symmetries. Possible existence of higher spin states 
with masses of the order of the Hubble parameter was investigated in Ref. \cite{Baumann} with the help of the  EFT. 
In search for heavy particle signature in the cosmic microwave background (CMB) data, 
it is prerequisite to know the 
standard model signals, which have been studied in Ref. \cite{SMcosmo}.

Motivated by these and related  developments \cite{achucarro, pisasaki, 
gwyn1, gong2, gwyn2, kehagias, dimastrogiovanni, 
chennamjoowang, meerburg},   
we apply in the present  paper,  the EFT method to the quasi-single field inflation model along the line similar 
to Noumi {\it  et al}'s  \cite{Noumi}, but paying more attention to tensor fluctuation $\gamma _{ij}$. 
Besides  the coupling between  $\zeta $ and $\sigma$ introduced in Ref. \cite{Noumi}, 
we study additionally the graviton coupling : `{\sl $\gamma\zeta\sigma$ coupling}'.
We compute the correlation functions 
\begin{eqnarray}
\langle \zeta \zeta \rangle , 
\qquad
\langle \gamma ^{s} \zeta \zeta \rangle , 
\qquad 
\langle \gamma ^{s_1} \gamma ^{s_2} \zeta \zeta \rangle  
\label{eq:012graviton}
\end{eqnarray}
due to  this new coupling for the case of 
 the soft-graviton.  Here,  superscripts $s$, $s_1$ and $s_2$ in $\gamma $'s denote polarization of the gravitons. 
These correlations will be expressed as functions of $c_{1}$, $c_{4}$ and $m$, 
where $c_1$ and $c_4$ are some constants that are introduced in the EFT, and  $m$ is the mass of $\sigma$. 
(The explicit forms of correlations due to these couplings  will be  shown in Section  \ref{section3} 
from (\ref{f1ori})
 through (\ref{f3muzui})).
If the future observation could tell us something about   these three correlation functions, then 
we can determine the three unknowns, i.e., $c_{1}$, $c_{4}$ and $m$. It could be possible hopefully 
 that the mass $m$ would turn 
out to be on the order of $10^{13} - 10^{14}$ GeV.

We then go on to generalize our calculation of the zero- , one- and two-graviton  correlation functions in (\ref{eq:012graviton})  to general $N$-graviton correlations. 
Namely we construct the coupling of $\zeta$, $\sigma$ and $N$ soft-gravitons and 
compute correlation functions 
\begin{equation}
\langle\gamma^{s_1}\cdots\gamma^{s_{N}}\zeta\zeta\rangle , 
\qquad 
\langle\gamma^{s_1}\cdots \cdots \gamma^{s_{2N}}\zeta\zeta\rangle 
\label{eq:Ngravitoncorrelation}
\end{equation}
by taking into account of  this coupling once and twice, respectively. 
Here we are assuming tacitly that this new $\zeta $, $\sigma$, $N$-graviton coupling is most dominant. 
By plotting these correlations  as  functions of $m$ for several $N$'s, we examine how the number of soft-gravitons affects the correlation function.
%Finally, we derive a relation, which relates $\langle\gamma^{s_1}\cdots\gamma^{s_{N+1}}\zeta\zeta\rangle$ 
%to $\langle\gamma^{s_1}\cdots\gamma^{s_{N}}\zeta\zeta\rangle$ in the $m\to\infty$ limit in this model, 
%and confirm that our results are consistent with the relation numerically for $m \gg H$.\vspace{1mm}

This paper is organized as follows. In Section  \ref{section2},  after  the quasi-single field inflation is reviewed, we  
extend it  by applying the EFT method, and thereby  
introducing the  $\zeta \sigma $ and $\gamma \zeta \sigma$  couplings. 
In Section \ref{section3} we compute the effects due to these new couplings on 
the correlation functions (\ref{eq:012graviton}) and examine their  $m$  dependence. 
Although the $m$-effects on  the correlations go down quickly as $m \to \infty$, 
the ratios of the correlation functions are shown in Section \ref{sec:largemassbehavior}
 to  approach  some definite numbers. 
The generalization to the $N$ graviton correlations (\ref{eq:Ngravitoncorrelation}) is given  in 
Section \ref{section5}.  Section  \ref{sectionsummary} is devoted to summary of the present work. 
Derivations of most of the 
integration formulae are relegated to Appendices.
Main results  of the present paper have  been reported by one of the authors 
in Ref. \cite{RSaito}.  
Throughout the present work, we set $c=\hbar=8\pi G=1$.

%%%%%%%%%%%%%%%%%%%%%%%%%%%%%%%%%%%%
\section{Preliminaries}
\label{section2}

In the following we will use the Einstein-Hilbert action for the gravitational part of the action and consider 
the Friedmann-Lemaitre-Robertson-Walker (FLRW) metric 
\begin{eqnarray} 
ds^2&=&-dt^2+a^2(t) d\vector{x}^2
=
a^2(t)(-d\eta^2+d\vector{x}^2).
\end{eqnarray}
as the classical background. 
Note that   $a(t)$ is the scale factor and $\eta$ is the conformal time connected with the 
coordinate time $t$ by   $d\eta \equiv dt/a(t)$.   
The   Hubble and  slow-roll parameters are defined as usual by 
 \begin{eqnarray}
H\equiv \frac{{\dot a}(t)}{a(t)}\:, 
\qquad
 \epsilon\equiv-\frac{\dot{H}}{H^2}. 
%\qquad
%\delta\equiv\frac{1}{2}\frac{\ddot{H}}{H\dot{H}}
 \end{eqnarray}

%%%%%%%%%%%%%%%%
\subsection{Quasi-single field inflation}

In  the quasi-single field inflation \cite{Quasi}, one introduces two kinds of scalar fields, namely, 
inflaton and massive isocurvaton fields. The inflaton field moves along the tangential direction of the turning trajectories in the space of scalar fields, and the isocurvaton field goes in the orthogonal direction. 
The Lagrangian we deal with is 
\begin{equation}
\begin{split}
\label{Quasiaction}
S_{m}
=
\int d^{4}x \sqrt{-g} \: {\cal L}_{m}
\\
 =
\int d^4 x\sqrt{-g}\biggl[-\frac{1}{2}&(\tilde{R}+\chi)^2 g^{\mu\nu}
\partial_{\mu}\theta\partial_{\nu}\theta
-\frac{1}{2}g^{\mu\nu}\partial_{\mu}\chi \partial_{\nu}\chi -V_{\textrm{sr}}(\theta)-V(\chi )\biggl],
\end{split}
\end{equation}
where $\theta$ and $\chi $ describe tangential  and radial directions, respectively 
and ${\tilde R}$ is a constant.
The usual slow-roll inflaton potential is denoted by  $V_{\textrm{sr}}(\theta)$ and $V(\chi )$ is a  
potential of the other scalar field $\chi $ and traps $\chi $ around some point. 
Our metric signature is $(-1,+1,+1,+1)$ for the flat Minkowski case.

%%%%%%%%%%%%%
%\begin{figure}[H]
%  \centering
%  \includegraphics[width=7cm]{quasifigure0.pdf}
%  \caption{The polar coordinates in the quasi-single field inflation}
%  \label{quasifigure}
%\end{figure}
%%%%%%%%%%%%%%

\begin{figure}[H]
  \centering
  \includegraphics[width=7cm]{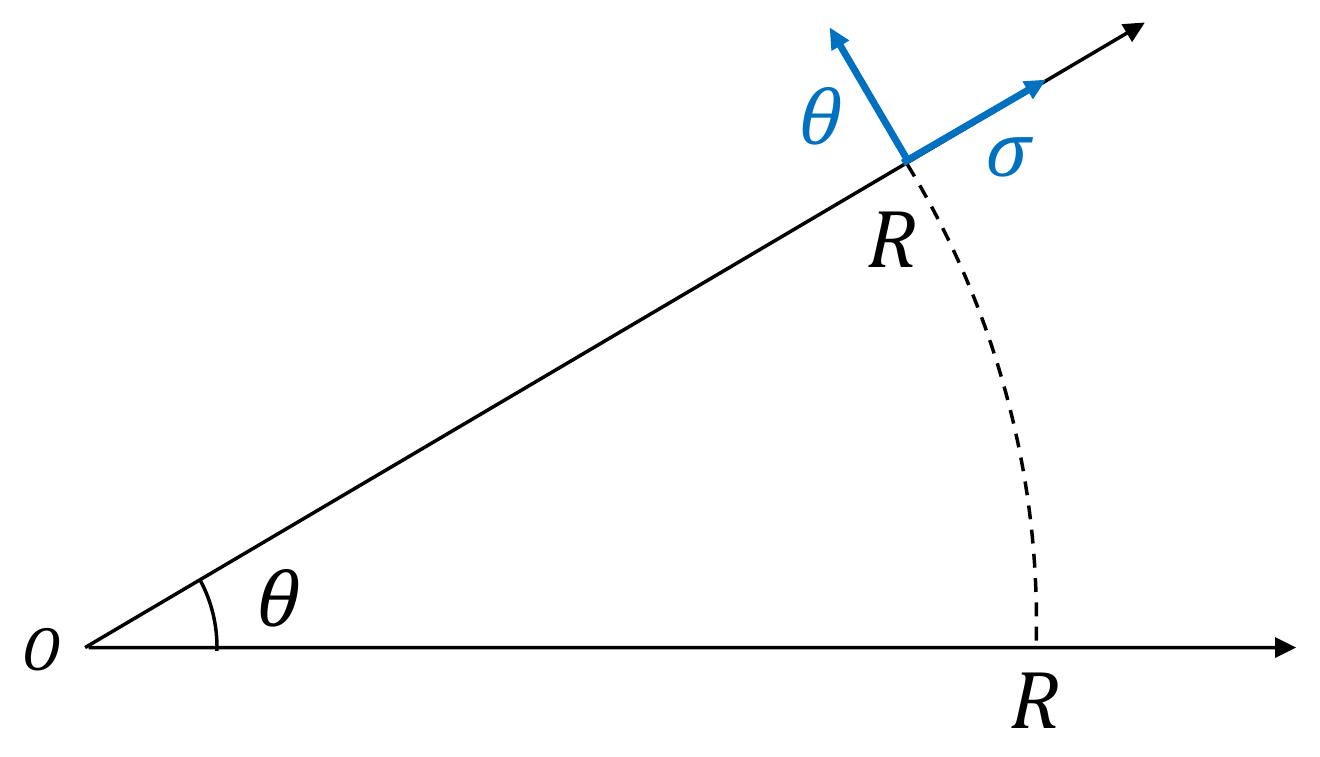}
  \caption{The polar coordinates in the field-space of the quasi-single field inflation. }
  \label{quasifigure}
\end{figure}

First of all, 
since we are working with FLRW classical background, we look for homogeneous and isotropic 
classical solutions for the scalar fields. 
Setting $\theta=\theta_0(t)$ and 
$\chi =\chi _0$ (constant), and using the action (\ref{Quasiaction}) 
to find the energy-momentum tensor in  the Einstein's equations, we obtain
\begin{eqnarray}
3H^2&=&\frac{1}{2}R^2\dot{\theta}_0^2+V(\chi_0)+V_{\textrm{sr}}(\theta_0),\label{relationQuasione}\\
-2\dot{H}&=&R^2\dot{\theta}_0^2\label{Quasihdot},\label{relationQuasitwo}
\end{eqnarray}
where $R\equiv\tilde{R}+\chi_0$. Then by  using the action (\ref{Quasiaction}) to derive 
the equations of motion (EOM) for $\theta = \theta_0(t)$ and $\chi = \chi_0 $, we arrive at 
\begin{eqnarray}
R^2\ddot{\theta}_0+3HR^2\dot{\theta}_0+V'_{\textrm{sr}}(\theta_0)&=&0,\\
R\dot{\theta}_0^2-V'(\chi_0)&=&0.\label{relationQuasithree}
\end{eqnarray}
Secondly, let us consider the fluctuations around the classical solutions of $\chi $ and the inflaton field $\theta$. 
In so doing, we use `{\sl the uniform inflaton gauge}':
\begin{equation}
\begin{split}
\label{uniforminflaton}
&\theta=\theta_0(t), \quad\chi=\chi_0+ \sigma(t,\vx),
%\\
%&h_{ij}=a^2(t)e^{2\zeta} \left [ \delta_{ij}  + \gamma _{ij} +\frac{1}{2} \gamma _{il} \gamma_{lj} + \cdots \right ] ,
\end{split}
\end{equation}
where the quantum fluctuation of $\theta $ is absent. 
  The geometrical picture of $R$,  $\theta $ and $\sigma (t, \vx) $ is illustrated  in 
Figure \ref{quasifigure}.
%%%%%%%%%%%%%
%%%%%%%%%%%%%%

Let us now decompose the metric through  the ADM formalism \cite{ADMpaper}  as 
\begin{equation}\label{ADMmetric}
ds^2=-N^2 dt^2+h_{ij}(dx^{i}+N^{i}dt)(dx^{j}+N^{j}dt).
\end{equation}
Note that $h_{ij}$ is the spatial metric 
\begin{equation}
\begin{split}
&h_{ij}=a^2(t)e^{2\zeta} \left [ \delta_{ij}  + \gamma _{ij} +\frac{1}{2} \gamma _{il} \gamma_{lj} + \cdots \right ] ,
\end{split}
\label{eq:hij}
\end{equation}
and that $N$ and $N^{i}$ are  called `lapse' and `shift', respectively. 
The tensor fluctuation $\gamma _{ij}$ in (\ref{eq:hij}) obeys transverse  and traceless conditions, i.e., 
$\partial _{i}\gamma _{ij}=0, \gamma _{ii}=0$.
The Einstein-Hilbert action  together with the matter part (\ref{Quasiaction}) in this ADM language 
 turns out to be 
\begin{equation}\label{QuasiADMaction}
S=\frac{1}{2}\int d^4 x\sqrt{h}\left[N\left(R^{(3)}+2\mathcal{L}_{m}\right)+N^{-1}\left(E_{ij}E^{ij}-E^2\right)\right],
\end{equation}
where $R^{(3)}$ is the  spatial scalar curvature and
\begin{eqnarray}
E_{ij}&\equiv&\frac{1}{2}(\dot{h}_{ij}-\nabla_{i}N_{j}-\nabla_{j}N_{i}),\\
E&\equiv&E^{i}_{i}.
\end{eqnarray}\vspace{3mm}
Note that ${\cal L}_{m}$ has been defined in  
(\ref{Quasiaction}). 
Then, setting $N=:1+N_1$ and solving the constraint equations
\begin{equation}\label{constraints}
\frac{\delta S}{\delta N}=0,\quad\frac{\delta S}{\delta N^{j}}=0, 
\end{equation}
to the first order, we obtain
\begin{eqnarray}
N_1&=&\frac{\dot{\zeta}}{H},\label{Quasiconstone}\\
\parti N^{i}&=&-a^{-2}\frac{1}{H}\partial^2\zeta+\epsilon\left(\zetadot-\frac{2H}{R}\sigma\right).\label{Quasiconsttwo}
\end{eqnarray}
The slow-roll parameter $\epsilon $ becomes 
$\epsilon\equiv-\dot{H}/H^2=R^2\dot{\theta}_0^2/2H^2$, according to 
 (\ref{Quasihdot}).

Substituting (\ref{Quasiconstone}) and (\ref{Quasiconsttwo}) into the action (\ref{QuasiADMaction}) and performing some integrations by parts, we arrive at 
\begin{eqnarray}
S_2
&\equiv &
S_{\zeta \zeta } + S_{\gamma \gamma} + S_{\sigma \sigma }
%\nonumber \\
%& & 
%+ \int d^4 x\biggl[
%\frac{1}{2}\left\{a^3(t)(\dot{\sigma})^2-a(t)(\partial\sigma)^2-a^3(t)
%\left(V''(\chi_0)-\dot{\theta}_0^2\right)(\sigma)^2\right\}
%\nonumber 
%\\
%& &
+ \int d^{4} x \biggl[ 
-2a^3(t)\frac{R\dot{\theta}_0^2}{H}\dot{\zeta}\,\sigma\biggl].
\label{eq:quadraticpartoftheaction}
\end{eqnarray}
for the quadratic part of the action, where 
%the quadratic part with respect to $\zeta $,  $\gamma $ and $\sigma $ are given by 
\begin{eqnarray}
S_{\zeta\zeta}
&=&
\int d^{4}x  \epsilon\left[a^{3}(t)\dot{\zeta}^{2}-a(t)(\partial\zeta)^{2}\right], 
\label{quadaction}
\\
S_{\gamma\gamma}
&=&
\frac{1}{8}\int d^{4}x\left[a^{3}(t)\dot{\gamma}_{ij}\dot{\gamma}_{ij}-a(t)\partial_{l}\gamma_{ij}\partial_{l}\gamma_{ij}\right], 
\label{quadactiongamma}
\\
S_{\sigma \sigma }
&=&
 \int d^4 x\biggl[
\frac{1}{2} a^3(t)(\dot{\sigma})^2-\frac{1}{2} a(t)(\partial\sigma)^2-
\frac{1}{2} a^3(t)\left(V''(\chi_0)-\dot{\theta}_0^2\right)(\sigma)^2\biggr]. 
\label{eq:ssigmasigma} 
\end{eqnarray}
The third  term in (\ref{eq:ssigmasigma})  is identified  as the mass term of $\sigma $, i.e., 
\begin{equation}
m^2\equiv V''(\chi_0)-\dot{\theta}_0^2\: .
\label{eq:massgivenbypotential}
\end{equation}
The fourth term in  (\ref{eq:quadraticpartoftheaction}) is a mixing interaction 
between $\zeta$ and $\sigma$,  giving rise to corrections to 
the power spectrum of $\zeta$.
Eqs. (\ref{quadaction}) and (\ref{quadactiongamma}) indicate that the propagation velocities of scalar and tensor perturbations are both 
unity in the current treatment.

%%%%%%%%%%%%%%%%%%%%%%%%%%%%%%%%
\subsection{Effective field theory approach }

Now that we have explained the quasi-single field inflation of Ref. \cite{Quasi}, 
we would like to deal with it in a more general way by applying ideas of the EFT.
Following the method of Ref. \cite{Effective}, 
we start from the unitary gauge and then employ the St{\" u}ckelberg trick,  i.e., 
translation in time direction, 
\begin{equation}
t \to {\tilde t}, \qquad \qquad t= {\tilde t} + \pi ( {\tilde t}, \vx),
\label{eq:fromttottilde}
\end{equation}
while the spatial coordinates $x^{i} $ are unchanged. 
The contravariant components of the metric transform under (\ref{eq:fromttottilde}) as 
\begin{eqnarray}
g^{\rho \sigma}&=&\frac{\partial x^{\rho}}{\partial\tilde{x}^{\mu}}\frac{\partial x^{\sigma}}{\partial\tilde{x}^{\nu}}\tilde{g}^{\mu\nu}\: , 
\end{eqnarray}
or more explicitly
\begin{eqnarray}
\label{eftg00}
g^{00}&=&(1+\pidot)^2\gtil^{00}+2(1+\pidot)(\parti\pi)\gtil^{0i}+(\parti\pi)(\partj\pi)\gtil^{ij}\:, 
\\
g^{0i}
&=&{\tilde g}^{0i}+ {\dot \pi} {\tilde g}^{0i} + (\partial _{j}\pi ) {\tilde g}^{ji}\:, 
\nonumber \\
g^{ij}&
=&{\tilde g}^{ij}\:, 
\nonumber 
\end{eqnarray}
where the dot means $d/d{\tilde t}$.
In the EFT, the $\pi $ field  is regarded as the Nambu-Goldstone mode and one can go over to the 
spatially flat gauge by this transformation.

The action that we are going to work with in the unitary gauge is given by 
\begin{equation}\label{eftaction10}
S=S_{0}+S_{\sigma}+S_{I},
\end{equation}
where
\begin{eqnarray}
S_{0}
&=&
\int d^4 x\sqrt{-g}\left[\frac{1}{2}R+\dot{H}(t)g^{00}-\left(3H^2(t)+\dot{H}(t)\right)\right],
\label{eftaction0}
\\
S_{\sigma}
&=&
\int d^4 x\sqrt{-g}\left[-\frac{1}{2}g^{\mu\nu}\partial_{\mu}\sigma\partial_{\nu}\sigma-\frac{1}{2}m^2 \sigma^2 +\cdots \right],
\label{sigmaaction}
\\
S_{I}
&=&
\int d^4 x\sqrt{-g}\left[c_1 \delta g^{00}\sigma+c_2 (\delta g^{00})^2 \sigma+c_3 \delta g^{00}\sigma^2 
+c_4 \delta g^{00}g^{\mu\nu}\partial_{\mu}\partial_{\nu}\sigma
\right].
\label{eftaction3}
\end{eqnarray}
Note that  $S_0$ is the standard part  in the EFT action.  
(The reader should not confuse the scalar curvature $R$ in (\ref{eftaction0}) with the radius 
$R={\tilde R}+\chi _{0}$ introduced previously.)
In principle we can think of terms containing  
\begin{equation}
\delta g^{00} \equiv g^{00}+1
\label{eq:deltagg+1}
\end{equation}
and higher powers thereof without $\sigma$ field in  (\ref{eftaction0}). 
We are, however, interested in  effects due to the  heavy fields  $\sigma$ 
in the present work, and so we do not include them in (\ref{eftaction0}). 
In (\ref{sigmaaction}) the mass of  $\sigma $ is denoted by $m$ and the ellipses 
$ \cdots $  correspond to terms coming from $\sigma$'s potential $V(\sigma)$.
The last one, $S_{I}$,  shows the interaction   of 
$\delta g^{00}$ and $\sigma$. 
The first three terms inside the pair of square brackets in (\ref{eftaction3}) will be seen to 
produce  terms which contain only $\zeta$ and $\sigma$ to the third order. 
In the unitary gauge, we could consider many more terms such as $\delta g^{00}\partial_{\mu}\sigma\partial^{\mu}\sigma$,   $(\delta g^{00})^2 \partial^0 \sigma$ and so on so 
forth,   but we will neglect these terms because 
they would contribute to the correlation functions (\ref{eq:012graviton}) only at the loop level.
The fourth term inside the pair of square brackets 
in  (\ref{eftaction3}) will be playing an  important role in the present work
because  it produces a $\gamma\zeta\sigma$ coupling 
with the graviton coming from $g^{\mu \nu}$. 
We assume that all of the coefficients 
$c_1$, $c_2$, $c_3$ and $c_4$ are position-independent constants but may in principle  
be time-dependent. In later calculations, however, we understand  that these coefficients are those 
at the time of horizon crossing. 
We could have included variations of the extrinsic curvature $K_{\mu \nu}$ in the most general setting. 
We,  however,  do not take those terms into account in the present work in order to avoid 
too much complication.

We are now in a position  to perform the time diffeomorphism 
(\ref{eq:fromttottilde}). Namely let us substitute 
$t=:\ttil+\pi(\ttil,\vx)$ and use (\ref{eftg00}) for $g^{00}$ and (\ref{eq:deltagg+1}) for $\delta g^{00}$. 
We then  solve the constraint equations as we did in (\ref{constraints}). 
To the first order, we can easily solve them and get (dropping all tildes):
\begin{eqnarray}
N_1&=&-\epsilon \left (-H\pi \right ),\\
\parti N^{i}&=&\epsilon\frac{d}{dt} \left (-H\pi \right )+c_1\frac{\sigma}{H}.
\label{eq:partialN}
\end{eqnarray}
We now see a $\sigma$-dependent term in (\ref{eq:partialN}). 
Then, substituting these solutions into the action (\ref{eftaction10}), and rewriting $\pi$ as 
\begin{equation}
\pi=-\frac{\zeta}{H} \: ,
\end{equation}
%which follows from (\ref{zetapi}) (dropping the subscript `$n$'), 
we can finally obtain the action in terms of  $\gamma$, $\zeta$ and $\sigma$.
The action for the part of $\zeta $ and $\gamma$ is the same as Maldacena's and 
the quadratic terms of the action,  in particular,  are
 given by $S_{\zeta\zeta}$  as in (\ref{quadaction}) and $S_{\gamma\gamma}$ as in (\ref{quadactiongamma}), respectively. 
Note that the velocities of scalar and tensor fluctuations both turn out to be unity in the present work. 
The quadratic part of the $\sigma $ field is also the same as (\ref{eq:ssigmasigma}) under the relation (\ref{eq:massgivenbypotential}), i.e., 
\begin{eqnarray}
S_{\sigma\sigma}&=&\int d^4 x\left[\frac{1}{2}a^3 (t)\sigdot^2-\frac{1}{2}a(t)(\partial\sigma)^2-\frac{1}{2}a^3(t)m^2\sigma^2 \right]\label{quadsigma}.
\end{eqnarray}
The new term including both $\zeta $ and $\sigma$ is
\begin{eqnarray}
S_{\zeta\sigma}&=&\int d^4 x\left[2\frac{c_1}{H}a^3(t)\zetadot\sigma\right].\label{quadzesig}
\end{eqnarray}
As for the cubic action, 
the expansion gives rise to a lot of terms involving combinations of  
$\zeta$ and $\sigma$ 
 such as $\zeta\sigma\sigma$ or $\zeta\zeta\sigma$, 
but as mentioned before, we will just examine a $\gamma\zeta\sigma$ coupling which is produced from the fourth term in (\ref{eftaction3}):
\begin{equation}\label{cubicgzsig}
S_{\gamma\zeta\sigma}=\int d^4 x\left[-2\frac{c_4}{H}a(t)\zetadot\gaij\parti\partj\sigma\right].
\end{equation}\vspace{2mm}

%%%%%%%%%%%%%%%%%%%%%%%%%%%%%%%%%%%
\subsection{Quasi-single field inflation versus EFT}

Let us check quickly how the set-up in the EFT  is related to the quasi-single field inflation.
We use again the uniform inflaton gauge (\ref{uniforminflaton}) in which the action 
(\ref{Quasiaction}) becomes
\begin{equation}
\begin{split}\label{actionrelationQuasi}
S_{m}=\int d^4 x\sqrt{-g}\biggl[-\frac{1}{2}&\left(R+ \sigma\right)^2 g^{00}\,\dot{\theta}_0^2\\
&-\frac{1}{2}g^{\mu\nu}\partial_{\mu}\sigma \partial_{\nu} \sigma -V_{\textrm{sr}}(\theta_0)-V(\chi_0 + \sigma)\biggl]. 
\end{split}
\end{equation}
%where $R\equiv\tilde{R}+\chi_0$.
Expanding the potential $V(\chi)$ as
\begin{equation} 
V(\chi_0 + \sigma)=V(\chi_0)+V'(\chi_0)\sigma+\frac{1}{2}V''(\chi_0) \sigma^{2} +\frac{1}{6}V'''(\chi_0) \sigma ^{3}+\cdots,
\end{equation}
and substituting (\ref{relationQuasione}), (\ref{relationQuasitwo}) and (\ref{relationQuasithree}) into (\ref{actionrelationQuasi}), we arrive at 
\begin{equation}
\begin{split}
S_{m}=\int &d^4 x\sqrt{-g}\biggl[\dot{H}(t)g^{00}-\left(3H^2(t)+\dot{H}(t)\right)\\
&-\frac{1}{2}g^{\mu\nu}\partial_{\mu} \sigma \partial_{\nu} \sigma 
-\frac{1}{2}\left(V''(\chi_0)-\dot{\theta}_0^2\right) \sigma^{2}-\frac{1}{6}V'''(\chi_0) \sigma ^{3}-\cdots\\
&\qquad\qquad-R\dot{\theta}_0^2\left(g^{00}+1\right) \sigma-\frac{1}{2}\dot{\theta}_0^2\left(g^{00}+1\right)\sigma^{2}\biggl].
\end{split}
\end{equation}
The first line corresponds to $S_0$ (\ref{eftaction0}) except for  the gravity term.
The second line corresponds to $S_\sigma$ (\ref{sigmaaction}), where $m^2\equiv V''(\chi_0)-\dot{\theta}_0^2$. %Note that $\sigma$ in the EFT action is equivalent with $\sigma$ in the quasi-single field inflation.
Finally the third line corresponds to $S_{I}$, i.e.,  (\ref{eftaction3}),  and 
in order to go back to the quasi-single field inflation 
we are led to set
\begin{equation}
c_1=-R\dot{\theta}_0^2,\quad c_2=0,\quad c_3=-\frac{1}{2}\dot{\theta}_0^2,\quad c_4=0.
\end{equation}
It follows therefore that the quasi-single field inflation belongs to the general class of inflationary models  
 offered  by the EFT method.

\vskip2mm
%{\color{red}{
In the present paper we would like to generalize the quasi-single field inflation model with the help of 
the EFT method by taking into account  the  $c_{1}$ and $c_{4}$ terms. The effects due to the $c_{2}$ terms 
on the other hand are  assumed to be  small and are simply neglected, which is in accordance with the original 
quasi-single field inflation.  
This implies that the diagram with $\zeta \zeta \sigma$ and $\gamma \gamma \sigma$ vertices is  
considered as subdominant.
With this assumption the scope of our analyses is kept within a reasonable size.  
The $c_{3}$ term gives rise to $\zeta \sigma \sigma$ type interactions and would contribute to the correlations 
(\ref{eq:012graviton}) and (\ref{eq:Ngravitoncorrelation}) only at higher orders containing $\sigma$-loops.
 For this reason the $c_{3}$ interactions are not considered in our work.
%}}

%%%%%%%%%%%%%%%%%%%%%%%%%%%%%%%%%%%%%%
\subsection{Quantization of the $\zeta$,  $\gamma$ and $\sigma$ fields}

Now we would like to quantize scalar and tensor fluctuations, $\zeta$ and  $\gamma$ 
together with the $\sigma$ field.  In the quantization procedure we will use 
 the conformal time $\eta$ defined by $d\eta \equiv dt/a(t)$. 
If we assume that $\eta$ moves from $-\infty$ to $0$ when $t$ moves from $-\infty$ to $\infty$, it follows that
\begin{equation}
\eta=-\frac{1}{aH}, 
\label{eq:eta=-1/aH}
\end{equation}
where we have ignored terms involving the slow-roll parameters. 
The vacuum is always assumed to be the Bunch-Davies vacuum.\vspace{3mm}

\noindent
\underline{Quantization of the $\zeta $ field}\vspace{3mm}

The equation of motion for $\zeta $ is seen from  the action (\ref{quadaction}) as 
\begin{equation}
\frac{d^{2} \zeta }{d\eta ^{2}}-\frac{2}{\eta} \frac{d\zeta}{d\eta} -\partial ^{2}\zeta =0. 
%\zeta''-\frac{2}{\eta}\zeta'-\partial^{2}\zeta=0,
\label{zetaeom}
\end{equation}
%where the prime $'$ means the differentiation $d/d\eta$. 
Note that we are taking the de Sitter limit, in which $H$ is almost constant and $\epsilon\ll 1$.
The way to quantize $\zeta$ is the same as the usual quantum field theoretical  method \cite{Birrel}. 
First of all,   we represent $\zeta$ using the Fourier transformation,
\begin{equation}
	\label{fourzeta}
	\color{black}
	\zeta(\eta, \boldsymbol x)
	=
	\int\frac{d^3k}{(2\pi)^3}
	\bigg[
		u_k(\eta)a_{\boldsymbol k}e^{i\boldsymbol k\cdot\boldsymbol x}
		+
		u_k^*(\eta)a_{\boldsymbol k}^\dagger e^{-i\boldsymbol k\cdot\boldsymbol x}
	\bigg]
\end{equation}
%%%%%%%%%%%%%%%%%%%%%%%%%%%%%%%%%%%%%%%%%
%\begin{equation}\label{fourzeta}
%\zeta(\eta,\vx)=\int\frac{d^3{k}}{(2\pi)^{3}}\left[u_{k}(\eta)a_{\scriptsize{\vk}}e^{i\vk\cdot\vx}+u_{k}^{*}
%(\eta)a_{\scriptsize{\vk}}^{\dagger}e^{-i\vk\cdot\vx}\right],
%\end{equation}
%%%%%%%%%%%%%%%%%%%%%%%%%%%%%%%%%%%%%%%%%%%%%%%%
where $u_{k}$ and $u_{k}^{*}$ are the solutions of EOM, 
  (\ref{zetaeom}) and $a_{\boldsymbol k}$ and $a_{\boldsymbol k}^{\dagger}$ are respectively the annihilation and creation operators.
They satisfy the commutation relation:
\begin{equation}\label{aacomm}
\left[a_{\boldsymbol k}, a_{\boldsymbol k'}^{\dagger}\right]=(2\pi)^{3}\delta^{3}(\boldsymbol k - \boldsymbol k').
\end{equation}
This equation together with  the canonical commutation relation
%%%%%%%%%%%%%%%%%%%%
%\begin{equation}
%\left[\zeta(\eta,\vx), \frac{\delta L}{\delta\zeta'}(\eta,\vy)\right]=i\delta^{3}(\vx-\vy),
%\end{equation}
%%%%%%%%%%%%%%%%%%
imposes the normalization conditions on $u_{k}$ and $u_{k}^{*}$, which turn out to be 
\begin{eqnarray}
u_{k}(\eta)&=&\frac{H}{\sqrt{4\epsilon k^3}}(1+i k\eta)e^{-i k\eta},\label{uksol}\\
u_{k}^{*}(\eta)&=&\frac{H}{\sqrt{4\epsilon k^3}}(1-i k\eta)e^{i k\eta}.\label{ukcsol}
\end{eqnarray}
%\vspace{3mm}
It is straightforward  to derive two-point function formulae, 
\begin{eqnarray}
\langle 0 \vert \zeta(t,\vx)\zeta(t',\vx') \vert 0 \rangle 
&=&
\int \frac{d^{3}k}{(2\pi)^{3}}e^{i{\boldsymbol k} \cdot(\boldsymbol x-\boldsymbol x')}\left[\frac{H^{2}}{4\epsilon k^{3}}(1+ik\eta)(1-i k\eta')e^{i k(\eta'-\eta)}\right], 
\label{eq:twopointfunctionzetazeta}
\\
\langle 0 \vert \zeta(t,\vx)\dot{\zeta}(t',\vx') \vert 0 \rangle 
&=&
\int \frac{d^{3}k}{(2\pi)^{3}}e^{i\boldsymbol k \cdot(\boldsymbol x -\boldsymbol x')}
\left[-\frac{H^{3}}{4\epsilon k}(1+ik\eta)\eta'^{2}e^{i k(\eta'-\eta)}\right], 
\label{eq:twopointfunctionzetazetadot}
\end{eqnarray}
which will be useful in our later computation.
Here $\vert 0 \rangle $ denotes the Bunch-Davis vacuum. 

%%%%%%%%%%%%%%%%%%%%%%%%%%%%%%%%%%%%
\vspace{3mm}
\noindent
\underline{Quantization of the $\gamma $ field}\vspace{3mm}

We can quantize the graviton field $\gamma_{ij}$ in a similar way. Let us recall that the quadratic action 
of $\gamma _{ij}$ is given by (\ref{quadactiongamma})
 and the EOM for $\gamma_{ij}$ turns out to be of the same form as (\ref{zetaeom}), 
i.e., 
\begin{eqnarray}
\frac{d^{2} \gamma_{ij}}{d\eta ^{2}}-\frac{2}{\eta} \frac{d\gamma_{ij}}{d\eta}
-\partial^{2}\gamma_{ij} =0.
\label{eq:EOMforgamma}
\end{eqnarray}
We now  represent $\gamma_{ij}$ in the Fourier transform as 
\begin{equation}\label{fourgamma}
\gamma_{ij}(\eta,\vx)=\sum_{s=+, \times}
\int\frac{d^3 k}{(2\pi)^3}\epsilon_{ij}^{s}(k)\left[U_{k}(\eta)b^{s}_{\boldsymbol  k}
e^{i\boldsymbol k \cdot \boldsymbol x}+U_{k}^{*}(\eta) b_{\boldsymbol k}^{s \dagger}e^{-i \boldsymbol k 
\cdot \boldsymbol x}\right],
\end{equation}
where $U_{k}$ and $U_{k}^{*}$ are the solutions of EOM (\ref{eq:EOMforgamma}) and are 
given by 
\begin{eqnarray}
U_{k}(\eta) & = & \frac{H}{\sqrt{k^{3}}} (1+ik\eta ) e^{-ik \eta }, 
\\
U_{k}^{*}(\eta ) & = &  \frac{H}{\sqrt{k^{3}}} (1 - ik\eta ) e^{ik \eta }, 
\end{eqnarray}
  and $\epsilon_{ij}^{s}$ is the polarization tensor  satisfying  the transverse and traceless condition, $k^{i}\epsilon_{ij}=\epsilon_{ii}=0$. 
The orthonormality condition is set as  $\epsilon_{ij}^{s}(k)\epsilon_{ij}^{s'}(k)=2\delta_{ss'}$. 
Also $b_{\boldsymbol k}^{s}$ and $b_{\boldsymbol k}^{s \dag}$are the annihilation and creation operators
 of graviton, respectively 
and satisfy the commutation relation
\begin{equation}
\left[ b_{\boldsymbol k}^{s}, b_{\boldsymbol k'}^{ s' \dagger}\right]=(2\pi)^{3}\delta^{3}(\boldsymbol k - 
\boldsymbol k')\delta _{s s'}.
\end{equation}
The two-point functions are easily worked out  as 
\begin{eqnarray}
\langle 0 \vert \gamma_{ij}(t,\vx)\gamma_{\alpha\beta}(t',\vx') \vert 0 \rangle 
&=&
\sum_{s}\int \frac{d^{3}q}{(2\pi)^{3}}e^{i\boldsymbol q \cdot(\boldsymbol x - \boldsymbol x')}\epsilon_{ij}^{s}\epsilon_{\alpha\beta}^{s}\left[\frac{H^{2}}{q^{3}}(1+iq\eta)(1-i q\eta')e^{i q(\eta'-\eta)}\right], 
\nonumber \\
\\
\langle 0 \vert \gamma_{ij}(t,\vx)\dot{\gamma}_{\alpha\beta}(t',\vx') \vert 0 \rangle 
&=&
\sum_{s}\int \frac{d^{3}q}{(2\pi)^{3}}e^{i \boldsymbol q \cdot (\boldsymbol x-\boldsymbol x')}\epsilon_{ij}^{s}\epsilon_{\alpha\beta}^{s}\left[-\frac{H^{3}}{q}(1+iq\eta)\eta'^{2}e^{i q(\eta'-\eta)}\right].
\nonumber \\
\end{eqnarray}

%%%%%%%%%%%%%%%%%%%%%%%%%%%%%%%55
\noindent
\underline{Quantization of the $\sigma $ field}\vspace{3mm}

The  quantization of $\sigma$ goes in the same way as $\zeta$. Firstly, we rewrite the quadratic 
action of $\sigma$ given in (\ref{quadsigma}) by using the conformal time  
(\ref{eq:eta=-1/aH}), and derive EOM
%%%%%%%%%%%%%%%%%%%%%%%%%%%%%%%%%%%%%%
%\begin{equation}\label{sigmaeom}
%\sigma''-\frac{2}{\eta}\sigma'-\partial^2 \sigma+\frac{m^2}{\eta^2 H^2}\sigma=0,
%\end{equation}
%%%%%%%%%%%%%%%%%%%%%%%%%%%%%%%%%%%%
\begin{equation}\label{sigmaeom}
\frac{d^{2}\sigma }{d\eta^{2}} -\frac{2}{\eta} \frac{d\sigma}{d\eta} -\partial ^{2} \sigma 
+\frac{m^{2}}{\eta^{2} H^{2} }\sigma =0.
\end{equation}
%where,  as mentioned before, the prime $'$ means differentiation with respect to the conformal time,  $d/d\eta$. 
Then we express  $\sigma$ in the Fourier transform as 
\begin{equation}
\label{sigmafour}
\sigma(\eta,\vx)=\int\frac{d^3{k}}{(2\pi)^{3}}\left[v_{k}(\eta)c_{\scriptsize{\vk}}e^{i \boldsymbol k \cdot 
\boldsymbol x }+v_{k}^{*}(\eta)c_{\scriptsize{\vk}}^{\dagger}e^{-i \boldsymbol k \cdot \boldsymbol x }
\right],
\end{equation}
where $v_{k}$ and $v_{k}^{*}$ are the solutions of EOM (\ref{sigmaeom}), and $c_{\boldsymbol k}$ 
and $c_{\boldsymbol k}^{\dagger}$ are the annihilation and creation operators  respectively that  satisfy 
the commutation relation given by  
\begin{equation}
\left[c_{\boldsymbol k }, c_{\boldsymbol k'}^{\dagger}\right]=(2\pi)^{3}\delta^{3}(\boldsymbol k-\boldsymbol k').
\end{equation}
 The solutions to (\ref{sigmaeom}) are already known 
(the reader is referred to Ref. \cite{Higuchi}  for example) to be 
\begin{eqnarray}
v_{k}(\eta)&=&-i e^{i(\nu+\frac{1}{2})\frac{\pi}{2}}\frac{\sqrt{\pi}}{2}H(-\eta)^{3/2}\hankel(-k\eta)\label{sigmavk},\\
v_{k}^{*}(\eta)&=&i e^{-i(\nu^{*}+\frac{1}{2})\frac{\pi}{2}}\frac{\sqrt{\pi}}{2}H(-\eta)^{3/2}\hankelc(-k\eta)\label{sigmavkc},
\end{eqnarray}
where
\begin{equation}
\nu\equiv\sqrt{\frac{9}{4}-\frac{m^2}{H^2}},
\end{equation}
and $\hankel$ is the Hankel function of the first kind. Using (\ref{sigmafour}) with (\ref{sigmavk}) and (\ref{sigmavkc}), we can compute the two point function of $\sigma$:
\begin{eqnarray}
& & \langle 0 \vert 
\sigma(\eta,\vx)\sigma(\eta',\vx')
\vert 0 \rangle 
\nonumber \\
& &\qquad  =\int \frac{d^{3}k}{(2\pi)^{3}}e^{i \boldsymbol k \cdot (\boldsymbol x -\boldsymbol x')} 
\left[e^{-\pi\textrm{Im}(\nu)}\frac{\pi}{4}H^2 (\eta\eta')^{\frac{3}{2}}\hankel(-k\eta)\hankelc(-k\eta')\right]
\label{tipssigma}
\end{eqnarray}

%%%%%%%%%%%%%%%%%%%%%%%%%%%%%%%%%%%%%%%%%%
\section{Computation of $\langle \zeta \zeta \rangle $, $\langle \gamma \zeta \zeta \rangle $, 
and $\langle \gamma \gamma \zeta \zeta \rangle $}
\label{section3}

\begin{figure}[H]
  \centering
  \includegraphics[width=7cm]{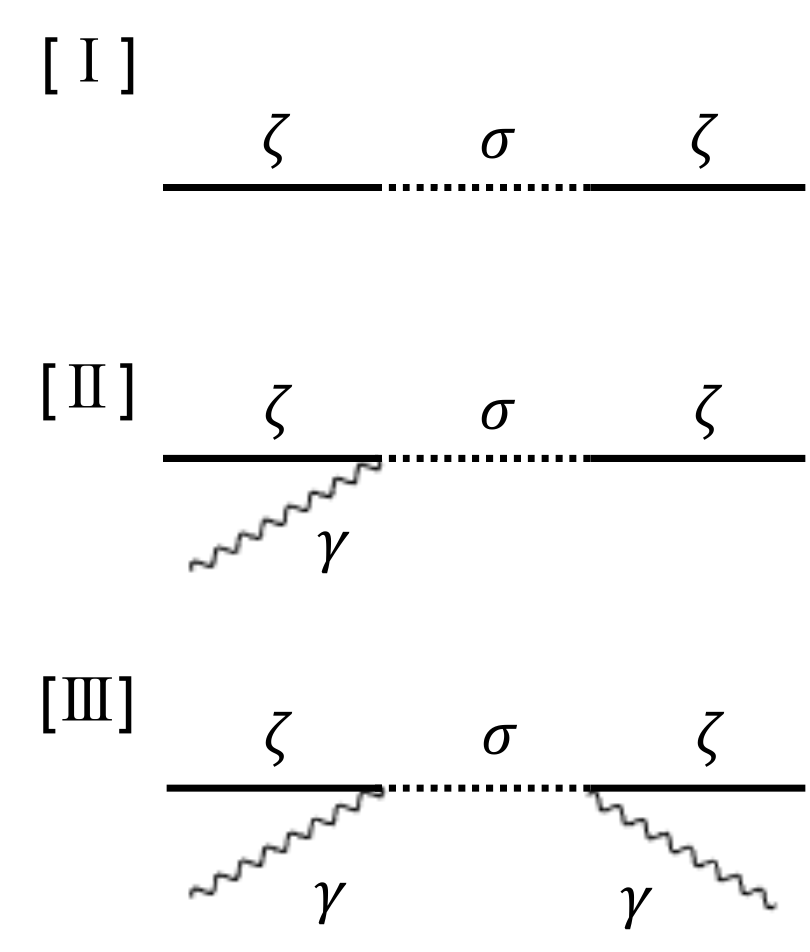}
  \caption{Effects due to the heavy field $\sigma $ on 
 [I] \,$\langle \zeta \zeta \rangle $, 
 [I\hspace{-.1em}I]  \,
 $ \langle \gamma \zeta \zeta \rangle $
 and 
 [I\hspace{-.1em}I\hspace{-.1em}I]\,
 $\langle \gamma \gamma \zeta \zeta \rangle $}
  \label{diagrams}
\end{figure}

We are now interested in the effects due to the interaction 
(\ref{eftaction3}), in particular due to the terms of $c_{1}$ and $c_{4}$. Obviously these terms affect the 
power spectrum $\langle \zeta \zeta \rangle $ and the 
correlations $\langle \gamma \zeta \zeta \rangle $, $\langle \gamma \gamma \zeta \zeta \rangle $. 
The relevant Feynman diagrams are  shown in Figure \ref{diagrams}.
%%%%%%%%%%%%%%%%%%%%%%%%
%%%%%%%%%%%%%%%%%%%%%%%%%%

\begin{figure}[H]
  \centering
  \includegraphics[width=7cm]{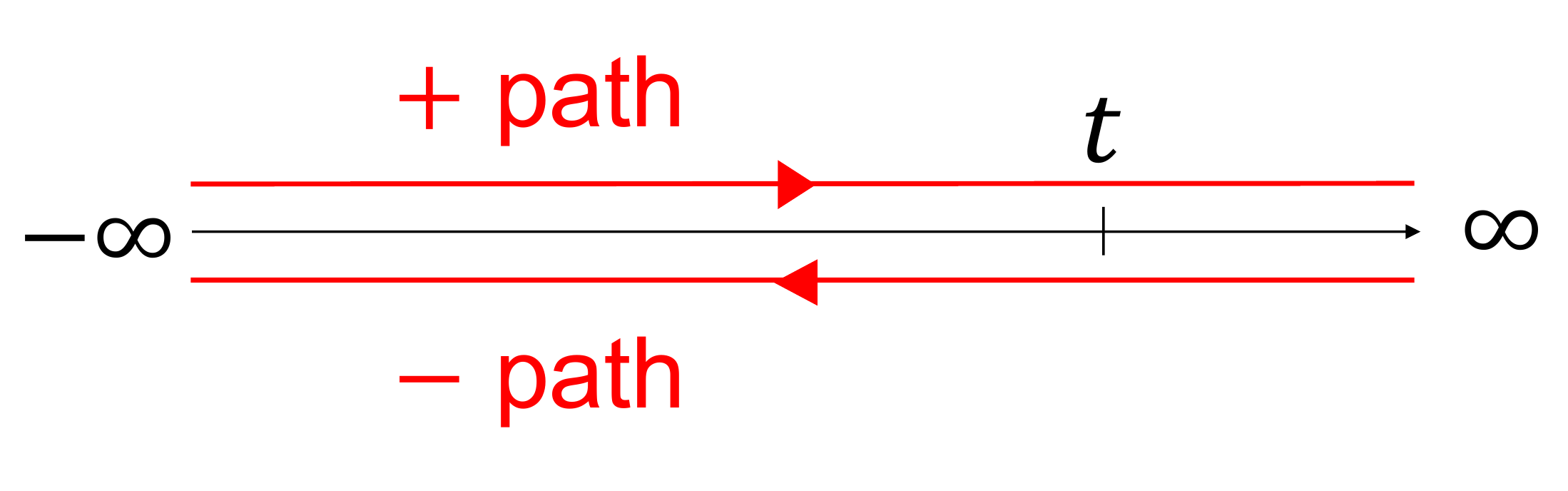}
  \caption{Time path in the in-in formalism. }
  \label{ininpath}
\end{figure}

The effect due to Figure \ref{diagrams}  [I] has  already been considered  by Chen and Wang in their quasi-single field model \cite{chenwang}. In the following, we would like to  generalize their method to compute also 
Figure \ref{diagrams}  [I\hspace{-.1em}I] and Figure \ref{diagrams} 
[I\hspace{-.1em}I\hspace{-.1em}I].
For this purpose we will use the in-in formalism \cite{ininpaper, keldysh, weinberg}, in which 
the expectation value of a certain operator 
$\mathcal{O}( t, \vx )$ can be computed by the formula 
\begin{eqnarray}
\langle\mathcal{O}(t,\vx)\rangle
&=&
\langle 0 \vert 
\overline{T} \left[\textrm{exp}\left[i\int_{-\infty}^{\infty}dt' H_{I}^{-}(t')\right]\right]T\left[\mathcal{O}^{+}(t,\vx)\textrm{exp}\left[-i\int_{-\infty}^{\infty}dt'H_{I}^{+}(t')\right]\right] \vert 0 \rangle  \:. 
\nonumber\\
\label{eq:expectationvalueO}
\end{eqnarray}
Here we label the interaction Hamiltonian $H_{I}$ and the operator $\mathcal{O}$ with `$+$' or `$-$' 
according to whether these operators are  on the `$+$ path' or `$-$ path' as shown in Figure \ref{ininpath}. 
Notice that on the $+$ path, operators are put in the usual time-ordering $(T)$, 
but on the $-$ path, operators are put in the anti-time ordering ($\overline{T}$).  
By shuffling Eq. (\ref{eq:expectationvalueO}), we are led to a more concise formula 
in the second order 
%%%%%%%%%%%%%%%%%%%%%%%
\begin{equation}
\begin{split}
&\langle  \mathcal{O}(t,\vx) \rangle \Big \vert _{{\rm 2nd. \; order}}
\\
&\qquad=\int_{-\infty}^{t}dt_1\int_{-\infty}^{t}dt_2 \langle 0 \vert H_{I}(t_1)\mathcal{O}(t,\vx)H_{I}(t_2) 
\vert 0 \rangle 
\\
&\qquad\qquad-2\textrm{Re}\left[\int_{-\infty}^{t}dt_1\int_{-\infty}^{t_1}dt_2 \langle 0 \vert 
\mathcal{O}(t,\vx)H_{I}(t_1)H_{I}(t_2) \vert 0 \rangle \right].
\end{split}\label{strategy}
\end{equation}
Here $\mathcal{O}(t, \vx ) $ is supposed to be either $\zeta \zeta $, $\gamma \zeta \zeta $ or $ \gamma \gamma \zeta \zeta $.
%%%%%%%%%%%%%%%%%%%%%%%%%%%%%%%%%%%%%%%%%%%
%%%%%%%%%%%%%%%%%%%%%%%%%%%%%%%%%%%%%%%%%%%%%

%%%%%%%%%%%%%%%%%%%%%%%%%%%%%%%%%%%%%%%%%%%%%%%
\subsection{Computation of $\langle \zeta \zeta \rangle $}

The interaction Hamiltonian $H_{I}(t)=-L_{I}(t)$ relevant to Figure \ref{diagrams} [I] is given 
by the action (\ref{quadzesig}):
\begin{equation}
H_{I}(t) \Big \vert _{{\rm Fig. } \: \ref{diagrams} \: [{\rm I} ]}
=-2\frac{c_1}{H}a^3 (t)\int d^3 x \,\zetadot\sigma.
%\label{eq.interactionhamiltonianI]
\end{equation}
Upon employing the two point functions 
(\ref{eq:twopointfunctionzetazeta}), 
(\ref{eq:twopointfunctionzetazetadot})
 and (\ref{tipssigma}), we find for (\ref{strategy})  with ${\cal O}=\zeta \zeta $
\begin{equation} 
\begin{split}
&\langle\zeta(\vk_1)\zeta(\vk_2)\rangle \Big\vert _{{\rm Fig.} \:  \ref{diagrams} \: [\rm{I} ]} \\
&=(2\pi)^3\delta^3 (\vk_1+\vk_2)\\
&\qquad\times\left(-2\frac{c_1}{H}\right)^2\left(\biarii\right)\left(\biari\right)\sigmatip\,
\times\,2!
\\
&\qquad\times\Biggl[\int_{-\infty}^{0}\frac{d\eta_1}{\eta_1^4H^4}\int_{-\infty}^{0}\frac{d\eta_2}{\eta_2^4H^4}\eta_1^2e^{-ik_1\eta_1}\eta_2^2e^{ik_2\eta_2}(\eta_1\eta_2)^{\frac{3}{2}}\hankel(-k_2\eta_1)\hankelc(-k_2\eta_2)\\
&\qquad-2\textrm{Re}\left[\int_{-\infty}^{0}\frac{d\eta_1}{\eta_1^4H^4}\int_{-\infty}^{\eta_1}\frac{d\eta_2}{\eta_2^4H^4}\eta_1^2e^{ik_1\eta_1}\eta_2^2e^{ik_2\eta_2}(\eta_1\eta_2)^{\frac{3}{2}}\hankel(-k_2\eta_1)\hankelc(-k_2\eta_2)\right]\Biggl].
\end{split}\label{zetzetmiddle}
\end{equation}
The  factor `$\times\,2!$'  in the third line in (\ref{zetzetmiddle}) 
is a combinatorial factor. 

We now evaluate (\ref{zetzetmiddle}) at $t\to\infty$, which corresponds to $\eta\to0$.
% for the diagram [I] in Figure \ref{diagrams}.\vspace{3mm}
If we set 
 $x\equiv-k_2\eta_1$ and $y\equiv-k_2\eta_2$,  then (\ref{zetzetmiddle}) becomes
\begin{equation}
\begin{split}
&\langle\zeta(\vk_1)\zeta(\vk_2)\rangle  \Big\vert _{{\rm Fig.} \:  \ref{diagrams} \: [\rm{I} ]} 
\\
&=(2\pi)^3\delta^3 (\vk_1+\vk_2)\ P_{\zeta}(k_2)\ \frac{\pi c_1^2}{2\epsilon H^4}e^{-\pi \textrm{Im}(\nu)}\\
&\qquad\times\Biggl[\left|\int_{0}^{\infty}dx\ x^{-\frac{1}{2}}e^{ix}\hankel(x)\right|^2\\
&\qquad\quad-2\textrm{Re}\left[\int_{0}^{\infty}dx\ x^{-\frac{1}{2}}e^{-ix}\hankel(x)\int_{x}^{\infty}dy\ y^{-\frac{1}{2}}e^{-iy}\hankelc(y)\right]\Biggl],
\end{split}
\label{zzresult}
\end{equation}
where  
\begin{equation}
P_{\zeta}(k)=\frac{H^2}{4\epsilon k^3} \:.
\end{equation}
would be  the power spectrum of the scalar perturbation in the absence of the heavy field $\sigma $.

\begin{figure}[H]
  \centering
  \includegraphics[width=9cm]{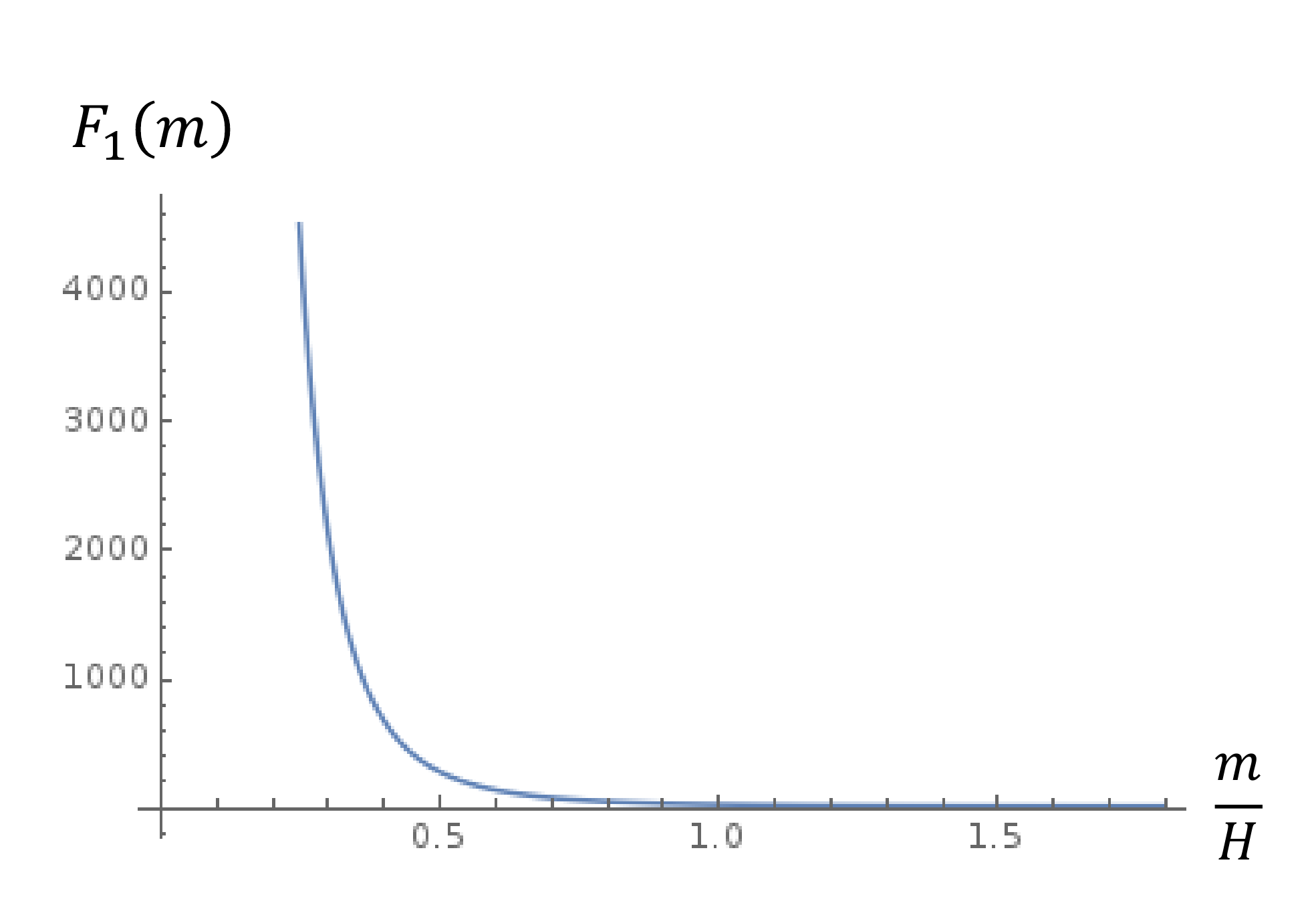}
  \caption{The plot for $F_1(m)$ with $m>0$}
  \label{fonem}
\end{figure}

The integrations in (\ref{zzresult}) are worked out in Appendices \ref{app:A} and \ref{app:B}
and in fact by putting 
(\ref{resulteasy}) 
and (\ref{eq:-1/2-1/2}) in  (\ref{zzresult}) we arrive at the following  formula
\begin{equation}
\begin{split}
\langle&\zeta(\vk_1)\zeta(\vk_2)\rangle  \Big\vert _{{\rm Fig. \: \ref{diagrams} \: (I) } }
=
(2\pi)^3\delta^3 (\vk_1+\vk_2)\,P_{\zeta}(k_2)\,\frac{c_1^2}{\epsilon H^4}F_1(m),
\end{split}\label{f1ori}
\end{equation}
where 
\begin{eqnarray}
F_1(m)
&\equiv &\frac{\pi^2}{\cosh^2(\pi\mu)}
-\frac{1}{\sinh(\pi\mu)}\textrm{Re}\left[\sum_{n=0}^{\infty}(-1)^{n}\Biggl\{\frac{e^{\pi\mu}}{(\frac{1}{2}+n+i\mu)^2}-\frac{e^{-\pi\mu}}{(\frac{1}{2}+n-i\mu)^2}\Biggl\}\right], 
\nonumber \\
\label{eq:f1mf1m}
\end{eqnarray}
with
\begin{eqnarray}
 \mu \equiv  \sqrt{\frac{m^{2}}{H^{2}}-\frac{9}{4} }=-i \nu \:.
\label{f1m}
\end{eqnarray}
In the course of deriving the formula (\ref{eq:f1mf1m}), we assumed that $\mu$ is real, namely, $m/H > 3/2$. 
The function $F_{1}(m)$, however, can be continued to the region $m/H < 3/2$ smoothly and its shape 
is as depicted in Figure \ref{fonem}. 
%The behavior of  $F_{1}(m)$ as a function of $m$ is depicted in Figure \ref{fonem}.
%%%%%%%%%%%%%%%%%%%%%%%%%%%
%%%%%%%%%%%%%%%%%%%%%%%%%%%%%
%%%%%%%%%%%%%%%%%%%%%%%%%%%
%\begin{figure}[H]
%  \centering
%  \includegraphics[width=9cm]{fonem.pdf}
%  \caption{The plot for $F_1(m)$ with $m>0$}
%  \label{fonem}
%\end{figure}
%%%%%%%%%%%%%%%%%%%%%%%%%%%%%
\noindent
Note that the function $F_{1}(m)$ has previously been evaluated numerically  by Chen and Wang \cite{chenwang} 
and our calculation is in good agreement with theirs.   An approximate analytic formula for $F_{1}(m)$ 
has also been studied in Ref. \cite{pisasaki}.

%%%%%%%%%%%%%%%%%%%%%%%%%%%%%%%%%%%%%%%%%%%%%%%%%
\subsection{Computation of $\langle \gamma \zeta \zeta \rangle $ in the soft graviton limit}

The interaction Hamiltonian relevant to Figure \ref{diagrams} [II]
consists of two terms, namely,  one from  (\ref{quadzesig}) and the other from (\ref{cubicgzsig}):
\begin{equation}
H_{I}(t)  \Big \vert _{{\rm Fig. } \: \ref{diagrams} \: [{\rm II } ]}
=-2\frac{c_1}{H}a^3 (t)\int d^3 x \,\zetadot\sigma+2\frac{c_4}{H}a(t)\int d^3 x \,\zetadot\gaij\parti\partj\sigma.
%\label{eq:interactionhamiltonianII}
\end{equation}
Using the two point functions (\ref{eq:twopointfunctionzetazeta}), (\ref{eq:twopointfunctionzetazetadot})
 and (\ref{tipssigma}), we find the formula 
 (\ref{strategy}) with $\mathcal{O}=\gamma \zeta \zeta $ turns out to be 
\begin{equation}
\begin{split}
&\langle\gamma^{s}(\vq)\zeta(\vk_1)\zeta(\vk_2)\rangle \Big\vert _{{\rm Fig. \: \ref{diagrams} \: [II] } }
\\
&=(2\pi)^3\delta^3 (\vq+\vk_1+\vk_2)\ \epsilon_{ij}^{s}(\vq)(-i)^2 (k_2)_{i}(k_2)_{j}\\
&\qquad\times\left(-4\frac{c_1 c_4}{H^2}\right)\binashiq\left(\biarii\right)\left(\biari\right)\sigmatip\,
\times\,2!
\\
&\qquad\times\Biggl[2\textrm{Re}\left[\int_{-\infty}^{0}\frac{d\eta_1}{\eta_1^4H^4}\int_{-\infty}^{0}\frac{d\eta_2}{\eta_2^2H^2}\eta_1^2e^{-ik_1\eta_1}\eta_2^2e^{ik_2\eta_2}(\eta_1\eta_2)^{\frac{3}{2}}\hankel(-k_2\eta_1)\hankelc(-k_2\eta_2)\right]\\
&\quad\qquad-2\textrm{Re}\left[\int_{-\infty}^{0}\frac{d\eta_1}{\eta_1^4H^4}\int_{-\infty}^{\eta_1}\frac{d\eta_2}{\eta_2^2H^2}\eta_1^2e^{ik_1\eta_1}\eta_2^2e^{ik_2\eta_2}(\eta_1\eta_2)^{\frac{3}{2}}\hankel(-k_2\eta_1)\hankelc(-k_2\eta_2)\right]\\
&\quad\qquad-2\textrm{Re}\left[\int_{-\infty}^{0}\frac{d\eta_1}{\eta_1^2H^2}\int_{-\infty}^{\eta_1}\frac{d\eta_2}{\eta_2^4H^4}\eta_1^2e^{ik_1\eta_1}\eta_2^2e^{ik_2\eta_2}(\eta_1\eta_2)^{\frac{3}{2}}\hankel(-k_2\eta_1)\hankelc(-k_2\eta_2)\right]\Biggl].
\end{split}\label{gzzmiddle}
\end{equation}
Note that we have already taken  the $\vq\to0$ limit to simplify our mathematical manipulation, and
 so $q$ does not appear in the above integrals.
The factor `$\times\,2!$' in the third line of (\ref{gzzmiddle})  is again a combinatorial factor.

By setting $x\equiv-k_2\eta_1$ and $y\equiv-k_2\eta_2$, (\ref{gzzmiddle}) may be transformed into 
\begin{equation}
\begin{split}
&\langle\gamma^{s}(\vq)\zeta(\vk_1)\zeta(\vk_2)\rangle  \Big\vert _{{\rm Fig. \: \ref{diagrams} \: [II] } }
\\
&=(2\pi)^3\delta^3 (\vq+\vk_1+\vk_2)\ \epsilon_{ij}^{s}(\vq)\frac{(k_2)_{i}(k_2)_{j}}{(k_2)^2}P_{\gamma}(q)P_{\zeta}(k_2) \frac{\pi c_1 c_4}{\epsilon H^2}e^{-\pi \textrm{Im}(\nu)}\\
&\qquad\times\Biggl[\textrm{Re}\left[\int_{0}^{\infty}dx\ x^{-\frac{1}{2}}e^{ix}\hankel(x)\int_{0}^{\infty}dy\ y^{\frac{3}{2}}e^{-iy}\hankelc(y)\right]\\
&\qquad\qquad-\textrm{Re}\left[\int_{0}^{\infty}dx\ x^{-\frac{1}{2}}e^{-ix}\hankel(x)\int_{x}^{\infty}dy\ y^{\frac{3}{2}}e^{-iy}\hankelc(y)\right]\\
&\qquad\qquad-\textrm{Re}\left[\int_{0}^{\infty}dx\ x^{\frac{3}{2}}e^{-ix}\hankel(x)\int_{x}^{\infty}dy\ y^{-\frac{1}{2}}e^{-iy}\hankelc(y)\right]\Biggl],
\end{split}\label{gzzresult}
\end{equation}
where 
%the power spectrum of the tensor perturbation is given by 
\begin{equation} 
P_{\gamma}(q)=\frac{H^2}{q^3}\:. 
\end{equation}
would be the power spectrum of the tensor perturbation in the  absence of $\sigma $. 
(Note in passing that this may differ from the formula often found in literatures by factor four.)

The reader is referred to Appendices \ref{app:A} and \ref{app:B} for the integration formulas for (\ref{gzzresult}). 
By combining  (\ref{resulteasy}), 
(\ref{resulteasy2}), (\ref{eq:-1/23/2}), and (\ref{eq:3/2-1/2}) altogether in (\ref{gzzresult})
one can easily find a formula
\begin{equation}
\begin{split}
\langle\gamma^{s}&(\vq)\zeta(\vk_1)\zeta(\vk_2)\rangle  \Big\vert _{{\rm Fig. \: \ref{diagrams} \: [II] } }
\\
&=(2\pi)^3\delta^3 (\vq+\vk_1+\vk_2)\ \epsilon_{ij}^{s}(\vq)\frac{(k_2)_{i}(k_2)_{j}}{(k_2)^2}P_{\gamma}(q)P_{\zeta}(k_2) \frac{c_1 c_4}{\epsilon H^2}\,F_2(m),
\end{split}\label{f2ori}
\end{equation}
where we have introduced the following function 
\begin{equation}
\begin{split}
F_2(m)\equiv&-\frac{1}{4}\left(\frac{1}{4}+\mu^2\right)\left(\frac{9}{4}+\mu^2\right)\frac{\pi^2}{\cosh^2(\pi\mu)}\\
&\quad\quad+\frac{1}{2\,\sinh(\pi\mu)}\,\textrm{Re}\Biggl[\,\sum_{n=0}^{\infty}(-1)^{n}(n+1)(n+2)\\
&\qquad\qquad\qquad\qquad\qquad\times\Biggl\{e^{\pi\mu}\frac{(1+n+2i\mu)(2+n+2i\mu)}{(\frac{1}{2}+n+i\mu)^2(\frac{5}{2}+n+i\mu)^2}\\
&\qquad\qquad\qquad\qquad\qquad\qquad-e^{-\pi\mu}\frac{(1+n-2i\mu)(2+n-2i\mu)}{(\frac{1}{2}+n-i\mu)^2(\frac{5}{2}+n-i\mu)^2}\Biggl\}\Biggl],
\end{split}\label{f2muzui}
\end{equation}
Recall that the quantity   $\mu $ has been introduced in (\ref{f1m}). 

\begin{figure}[H]
  \centering
  \includegraphics[width=9cm]{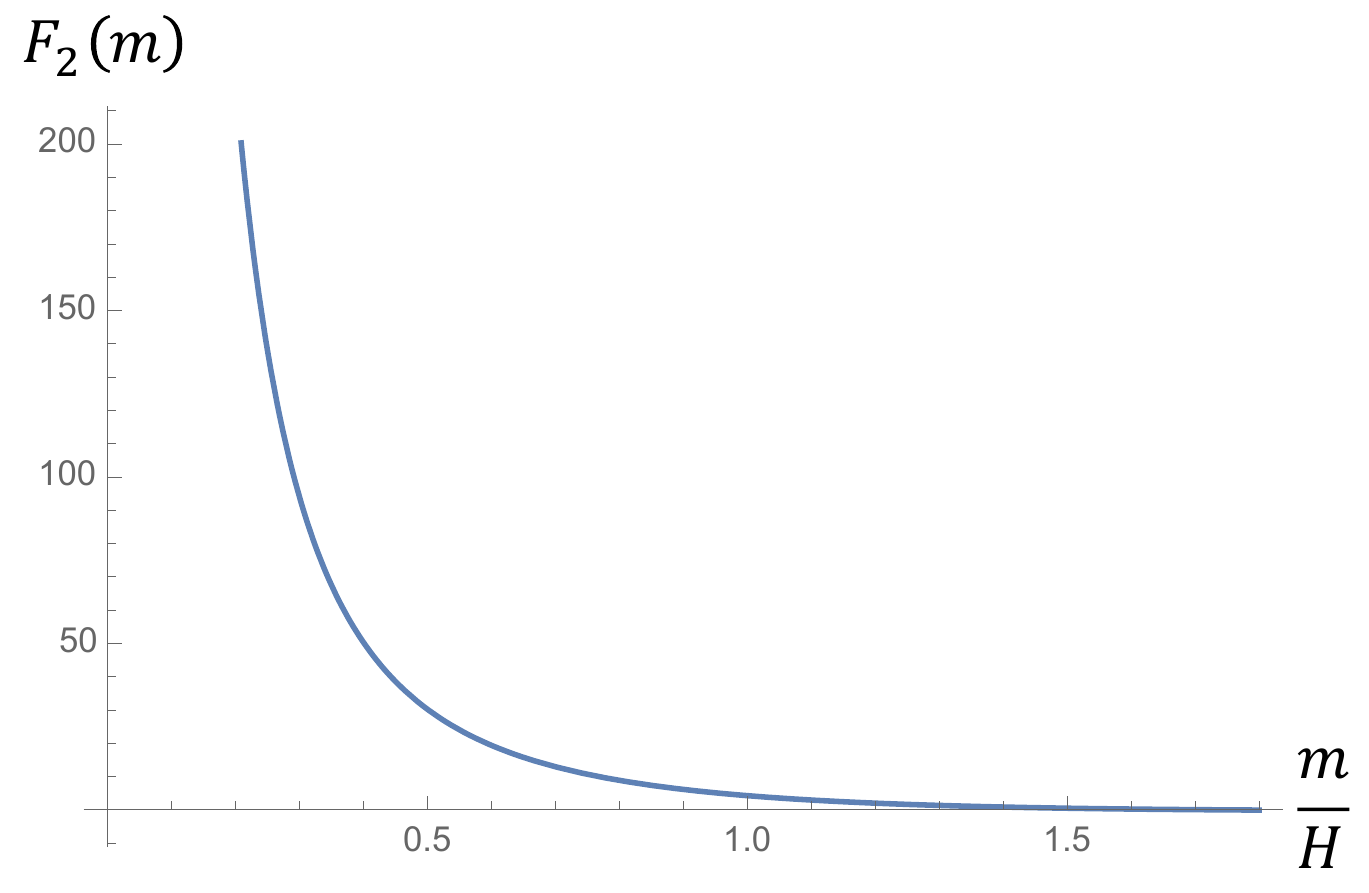}
  \caption{The plot for $F_2(m)$ with $m>0$}
  \label{ftwom}
\end{figure}

%{\color{blue}{
We have to pay a careful attention to the summation over $n$ in (\ref{f2muzui}). Because of the factor $(-1)^{n}$, it is an 
alternating sum, but the large $n$ behavior of each term is not mild enough to confirm the convergence.
The source of this indefinite nature of the summation may be traced back to the oscillatory behavior 
of the Hankel functions in (\ref{gzzresult}) for large $x$ and large $y$ and the factor $x^{3/2}$ and/or $y^{3/2}$.
 We may regularize the integrals in (\ref{gzzresult}) 
by deforming  the integration paths slightly away from the real axis on the complex $x$ and $y$ planes so that 
the integrals should converge. 
This type of regularization would lead us eventually to the regularization of the indefinite sum in (\ref{f2muzui}), 
such as zeta-function regularization. 
In the actual numerical calculation of $F_{2}(m)$, we assume that we are allowed to make use of the 
zeta-function regularization method in the summation in (\ref{f2muzui}). We will also employ the zeta-function regularization 
in subsequent calculations in the present paper.
%}}
The function $F_{2}(m)$ thus regularizaed is shown in Figure \ref{ftwom}, which  
looks as if it were decreasing monotonically as the mass $m$ becomes large. This is, however, not the case. 
In fact $F_{2}(m)$ has a local minimum at
$m/H \sim$ 2.3, which is outside in Figure \ref{ftwom}. We will come to this point later, when we illustrate  $F_{2}(m)$ 
in a different region of $m/H$ in Figure \ref{comparefs}. 
%%%%%%%%%%%%%%%%%%%%%%%%%%%%%%%%
%%%%%%%%%%%%%%%%%%%%%%%%%%%%%%%%%%%%%%%

%%%%%%%%%%%%%%%%%%%%%%%%%%%%%%%%%%%%%%%%%%%%
\subsection{Computation of $\langle \gamma \gamma \zeta \zeta \rangle $ in the soft graviton limit}

The interaction Hamiltonian that is necessary to evaluate 
Figure   \ref{diagrams} [III]
is given simply  by (\ref{cubicgzsig}):
\begin{equation}
H_{I}(t)  \Big \vert _{{\rm Fig. } \: \ref{diagrams} \: [{\rm III } ]}
=2\frac{c_4}{H}a(t)\int d^3 x \,\zetadot\gaij\parti\partj\sigma.
%\label{eq:interactionhamiltonianIII}
\end{equation}
A straightforward application of the two-point functions given previously 
provides us with   
the formula  (\ref{strategy}) with $\mathcal{O}=\gamma \gamma \zeta \zeta $  in the 
following way, 
\begin{equation}
\begin{split}
&\langle\gamma^{s_1}(\vq_1)\gamma^{s_2}(\vq_2)\zeta(\vk_1)\zeta(\vk_2)\rangle 
\Big\vert_{ {\rm Fig. \: \ref{diagrams} \: [III] }}
\\
&=(2\pi)^3\delta^3 (\vq_1+\vq_2+\vk_1+\vk_2)\ \epsilon_{ij}^{s_1}(\vq_1)(-i)^2 (k_1)_{i}(k_1)_{j}\ \epsilon_{\alpha\beta}^{s_2}(\vq_2)(-i)^2 (k_2)_{\alpha}(k_2)_{\beta}\\
&\qquad\times\left(2\frac{c_4}{H}\right)^2\frac{H^2}{(q_1)^3}\frac{H^2}{(q_2)^3}\left(\biarii\right)\left(\biari\right)\sigmatip\,
\times\,2!\times2!
\\
&\qquad\times\Biggl[\int_{-\infty}^{0}\frac{d\eta_1}{\eta_1^2H^2}\int_{-\infty}^{0}\frac{d\eta_2}{\eta_2^2H^2}\eta_1^2e^{-ik_1\eta_1}\eta_2^2e^{ik_2\eta_2}(\eta_1\eta_2)^{\frac{3}{2}}\hankel(-k_2\eta_1)\hankelc(-k_2\eta_2)\\
&\qquad-2\textrm{Re}\left[\int_{-\infty}^{0}\frac{d\eta_1}{\eta_1^2H^2}\int_{-\infty}^{\eta_1}\frac{d\eta_2}{\eta_2^2H^2}\eta_1^2e^{ik_1\eta_1}\eta_2^2e^{ik_2\eta_2}(\eta_1\eta_2)^{\frac{3}{2}}\hankel(-k_2\eta_1)\hankelc(-k_2\eta_2)\right]\Biggl].
\end{split}\label{ggzzmiddle}
\end{equation}
Note that we have already taken 
 the double-soft limit $\vq_1, \vq_2 \to 0$ for the sake of calculational simplicity.
The  factor `$\times\,2!\times2!$' in the third line of   (\ref{ggzzmiddle})  is, as before, 
 a combinatorial factor for 
 Figure \ref{diagrams}
[I\hspace{-.1em}I\hspace{-.1em}I] .

\begin{figure}[H]
  \centering
  \includegraphics[width=9cm]{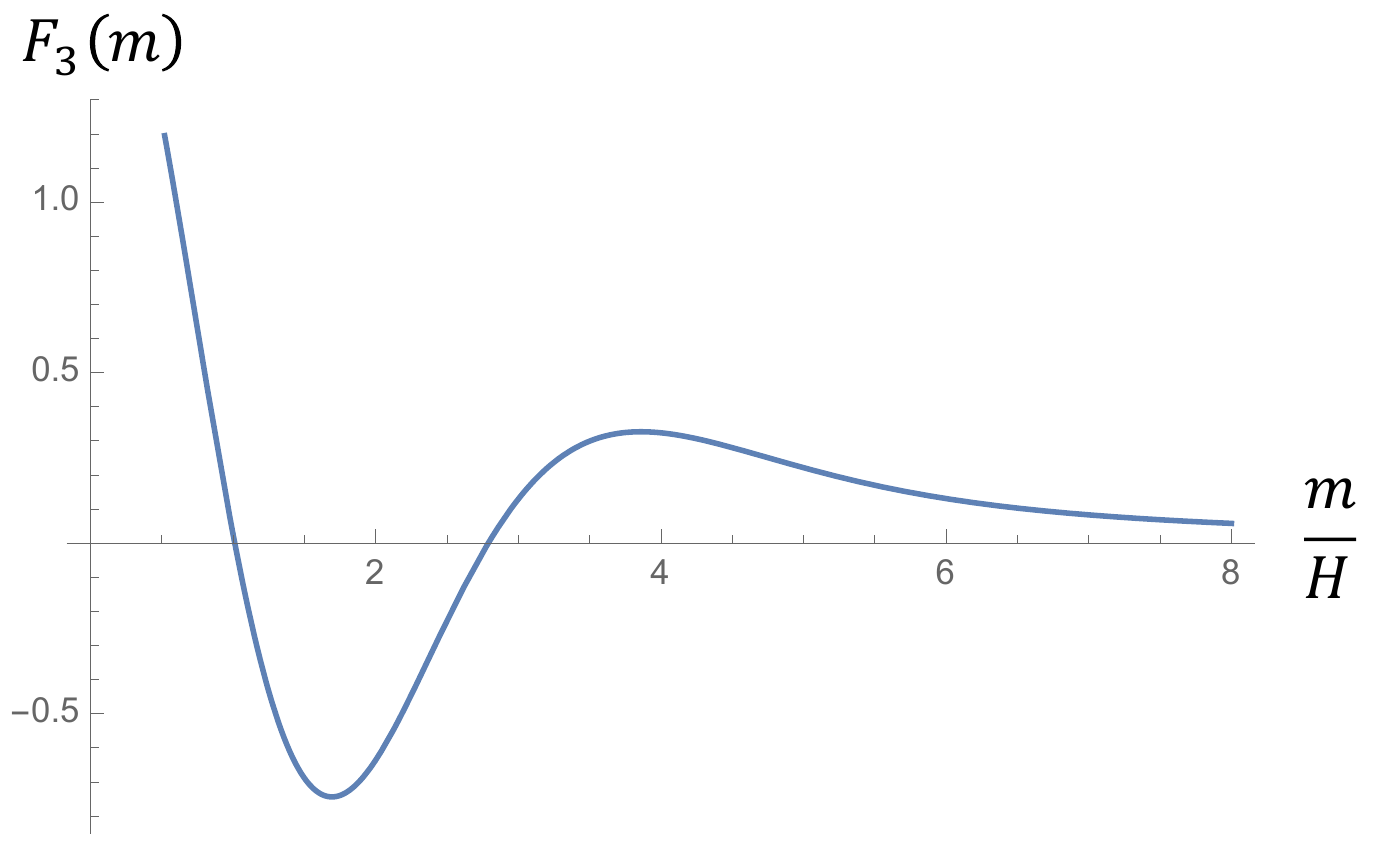}
  \caption{The plot for $F_3(m)$ in $m>0$}
  \label{fthreem}
\end{figure}

If we change the integration variables from $\eta_{1}$ and $\eta_{2}$ into 
$x\equiv-k_2\eta_1$ and $y\equiv-k_2\eta_2$ respectively,  then the correlation 
function (\ref{ggzzmiddle}) becomes
\begin{equation}
\begin{split}
&\langle\gamma^{s_1}(\vq_1)\gamma^{s_2}(\vq_2)\zeta(\vk_1)\zeta(\vk_2)\rangle
\Big\vert_{ {\rm Fig. \: \ref{diagrams} \: [III] }}
\\
&=(2\pi)^3\delta^3 (\vq_1+\vq_2+\vk_1+\vk_2)\\
&\qquad\times\epsilon_{ij}^{s_1}(\vq_1)\frac{(k_2)_{i}(k_2)_{j}}{(k_2)^2}\epsilon_{\alpha\beta}^{s_2}(\vq_2)\frac{(k_2)_{\alpha}(k_2)_{\beta}}{(k_2)^2}P_{\gamma}(q_1)P_{\gamma}(q_2)P_{\zeta}(k_2)\frac{\pi c_4^2}{\epsilon}e^{-\pi \textrm{Im}(\nu)}\\
&\qquad\times\Biggl[\left|\int_{0}^{\infty}dx\ x^{\frac{3}{2}}e^{ix}\hankel(x)\right|^2\\
&\qquad\qquad-2\textrm{Re}\left[\int_{0}^{\infty}dx\ x^{\frac{3}{2}}e^{-ix}\hankel(x)\int_{x}^{\infty}dy\ y^{\frac{3}{2}}e^{-iy}\hankelc(y)\right]\Biggl]. 
\end{split}\label{ggzzresult}
\end{equation}
In order to look at the $m$-dependence of  
 (\ref{ggzzresult}) more closely, 
we may put the  integration formulae in (\ref{resulteasy}) and (\ref{eq:3/23/2})
and then we end up with 
\begin{equation}
\begin{split}
&\langle\gamma^{s_1}(\vq_1)\gamma^{s_2}(\vq_2)\zeta(\vk_1)\zeta(\vk_2)\rangle
\Big\vert_{ {\rm Fig. \: \ref{diagrams} \: [III] }}
\\
&\quad=(2\pi)^3\delta^3 (\vq_1+\vq_2+\vk_1+\vk_2)\\
&\qquad\quad\times\epsilon_{ij}^{s_1}(\vq_1)\frac{(k_2)_{i}(k_2)_{j}}{(k_2)^2}\epsilon_{\alpha\beta}^{s_2}(\vq_2)\frac{(k_2)_{\alpha}(k_2)_{\beta}}{(k_2)^2}P_{\gamma}(q_1)P_{\gamma}(q_2)P_{\zeta}(k_2)\frac{c_4^2}{\epsilon}\,F_3(m),
\end{split}\label{f3ori}
\end{equation}
where the function $F_{3}(m)$ defined by  
\begin{equation}
\begin{split}
F_3&(m)\equiv\\
&\frac{1}{32}\left(\frac{1}{4}+\mu^2\right)^2\left(\frac{9}{4}+\mu^2\right)^2\frac{\pi^2}{\cosh^2(\pi\mu)}\\
&-\frac{1}{8\,\sinh(\pi\mu)}\\
&\quad\quad\times\textrm{Re}\Biggl[\,\sum_{n=0}^{\infty}(-1)^{n}(n+1)(n+2)(n+3)(n+4)\\
&\qquad\times\Biggl\{e^{\pi\mu}\frac{(1+n+2i\mu)(2+n+2i\mu)(3+n+2i\mu)(4+n+2i\mu)}{(\frac{1}{2}+n+i\mu)(\frac{3}{2}+n+i\mu)(\frac{5}{2}+n+i\mu)^2(\frac{7}{2}+n+i\mu)(\frac{9}{2}+n+i\mu)}\\
&\quad\qquad-e^{-\pi\mu}\frac{(1+n-2i\mu)(2+n-2i\mu)(3+n-2i\mu)(4+n-2i\mu)}{(\frac{1}{2}+n-i\mu)(\frac{3}{2}+n-i\mu)(\frac{5}{2}+n-i\mu)^2(\frac{7}{2}+n-i\mu)(\frac{9}{2}+n-i\mu)}\Biggl\}\Biggl].
\end{split}\label{f3muzui}
\end{equation}
has a peculiar behavior as is shown in Figure \ref{fthreem}. 
%%%%%%%%%%%%%%%%%%
%%%%%%%%%%%%%%%%%%%

%%%%%%%%%%%%%%%%%%%%%%%%%%%%%%%%%%%%%%%%%%%%%%
\section{The large mass limit $m \to \infty$}
\label{sec:largemassbehavior}

\begin{figure}[H]
  \centering
  \includegraphics[width=9cm]{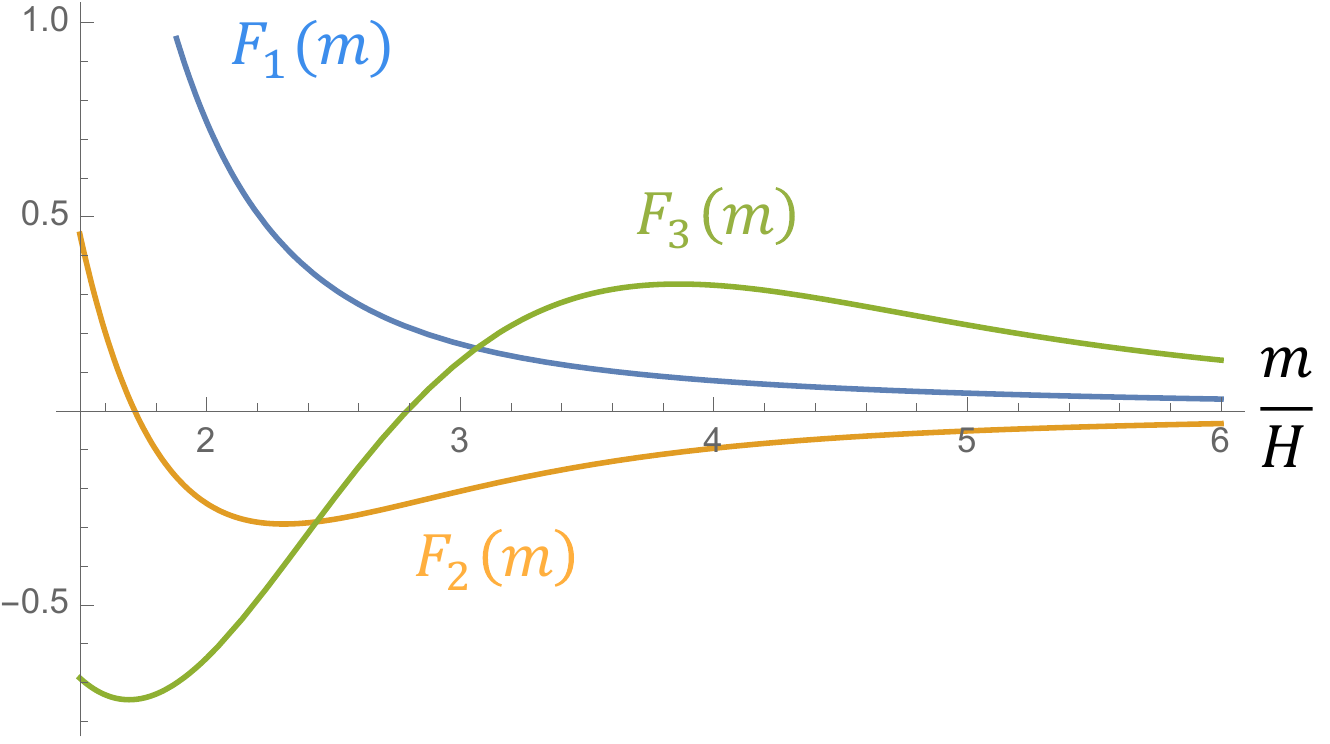}
  \caption{The plots for $F_1(m)$, $F_2(m)$ and $F_3(m)$ in $m\geq1.5H$}
  \label{comparefs}
\end{figure}

Since we are interested in the signature of the heavy field $\sigma $ in the correlation functions, 
we would like to scrutinize the $m$-dependence of the three characteristic 
functions $F_{i}(m)\:\:\:(i=1,2,3)$ 
which are drawn  together in  Figure \ref{comparefs}. 
Note that $F_{2}(m)$ has a local minimum around $m/H \sim 2.3$, which was not seen in Figure \ref{ftwom}.
While we can see local maxima and local minima in 
 Figure \ref{comparefs}, the three functions approach all to zero as 
$m \to \infty $. This is natural because 
heavy particle effects are in general expected to disappear at lower energy, 
 which is in the present case the Hubble parameter $H$.
%%%%%%%%%%%%%%%%%%%%%%%%%%%
%%%%%%%%%%%%%%%%%%%%%%%%%%%

Curiously, 
the rate by which $F_{1}(m)$, $F_{2}(m)$ and $F_{3}(m)$ approach the horizontal axis 
as $m \to \infty $ seems to be shared commonly 
by these three functions.
% $F_{i}(m)$. 
In order to examine this point further,  let us consider the ratios 
\begin{equation}
G_{12}(m)\equiv\frac{F_1(m)}{F_2(m)},\quad G_{23}(m)\equiv\frac{F_2(m)}{F_3(m)}.
\end{equation}
We have  performed   numerical analyses of these ratios  by Mathematica 11,  the result of which
is given in Table \ref{numericalresult}  below:
\begin{table}[H]
  \centering
  \caption{The numerical analysis of $G_{12}(m)$ and $G_{23}(m)$}
  \begin{tabular}{|c||c|c|c|c|c|c|}\hline
  $\displaystyle m/H$&$\displaystyle 5.0$&$\displaystyle 10.0$&$\displaystyle 15.0$&$\displaystyle 20.0$&$\displaystyle 25.0$&$\displaystyle 30.0$\\ \hline\hline
  $\displaystyle G_{12}(m)$&$\displaystyle -0.88031$&$\displaystyle -0.978412$&$\displaystyle -0.99082$&$\displaystyle -0.99491$&$\displaystyle -0.996764$&$\displaystyle -0.99776$\\ \hline
  $\displaystyle G_{23}(m)$&$\displaystyle -0.233967$&$\displaystyle -0.310469$&$\displaystyle -0.323965$&$\displaystyle -0.328174$&$\displaystyle -0.330123$&$\displaystyle -0.331035$\\ \hline
  \end{tabular}\label{numericalresult}
\end{table}

\noindent
This result suggests the limiting behavior 
\begin{equation}
\lim_{m \to \infty}  G_{12}(m)\to-1,  \qquad  \lim _{m \to \infty} G_{23}(m)\to- \frac{1}{3}.
\label{eq:asymptoticvalues}
\end{equation}
%\subsubsection{$m\to\infty$ limit}\label{mtoinftysec}
Looking at the expressions (\ref{f1m}), (\ref{f2muzui}) and (\ref{f3muzui}), we do not find it easy 
to work out explicitly the asymptotic behaviors of $F_{i}(m)\: (i=1,2,3)$ as $m \to \infty$.
We are, however, able to determine the asymptotic values 
(\ref{eq:asymptoticvalues}) by employing  an alternative method.

\begin{figure}[H]
  \centering
  \includegraphics[width=10cm]{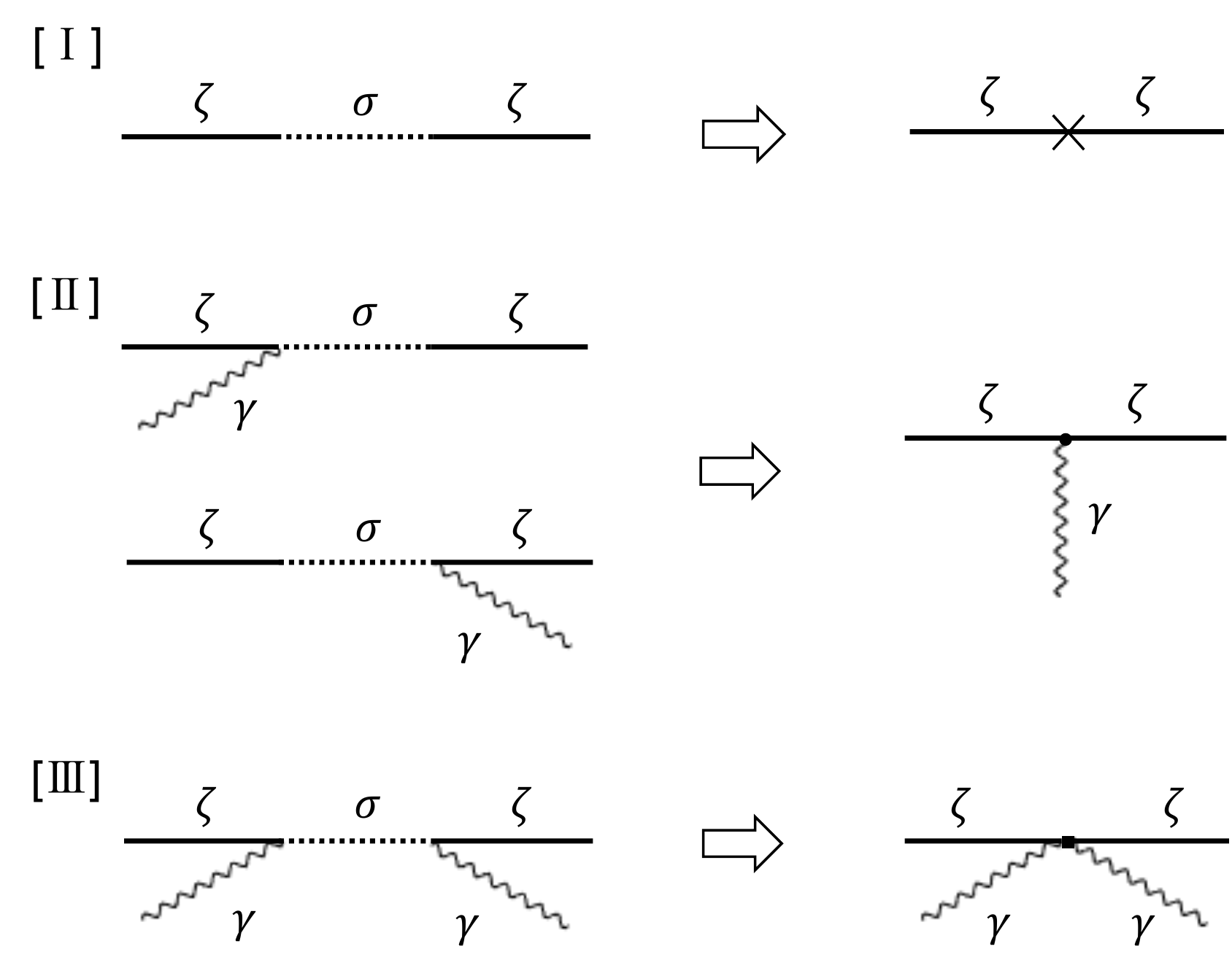}
  \caption{The original Feynman diagrams (left) are simplified into the right diagrams
 as  $m\to\infty$}
  \label{effectivefigure}
\end{figure}

Suppose that the mass of $\sigma $ is extremely large and the $\sigma $ field  is integrated out.
Then the interaction Hamiltonians 
 relevant to the three diagrams in Figure \ref{diagrams}
are reduced, by using (\ref{quadzesig}) and (\ref{cubicgzsig}),   to 
\begin{eqnarray}
 H_{I}(t)
 \Big\vert_{ {\rm Fig. } \: \ref{effectivefigure} \: [{\rm I}]}
&=&
\left(-2\frac{c_1}{H}\right)^2 a^3(t)f(m)\int d^3 x\ \zetadot^2,
\label{hamihami1}
\\
H_{I}(t) 
 \Big\vert_{ {\rm Fig. } \: \ref{effectivefigure} \: [{\rm II}]}
&=&\left(-2\frac{c_1}{H}\right)\left(2\frac{c_4}{H}\right)a(t)f(m)\int d^3 x\ 
\zetadot\,\gaij\parti\partj\zetadot\ \times2,
\label{hamihami2}
\\
H_{I}(t) 
 \Big\vert_{ {\rm Fig. } \: \ref{effectivefigure} \: [{\rm III}]}
&=&\left(2\frac{c_4}{H}\right)^2 a^{-1}(t)f(m)\int d^3 x\ \gaij\parti\partj\zetadot\,\gamma_{\alpha\beta}\partial_{\alpha}\partial_{\beta}\zetadot. 
\label{hamihami3}
\end{eqnarray}
This simplification is illustrated in Figure \ref{effectivefigure}.
Note that  $f(m)$ in (\ref{hamihami1}), (\ref{hamihami2}) and (\ref{hamihami3}) is 
a certain common function of $m$. 
Incidentally we have to  be careful about  the power  of $a(t)$ in  (\ref{hamihami1}),  (\ref{hamihami2}) and  (\ref{hamihami3}). 
The volume element of the integration of course gives us 
$\sqrt{-g}=a^3(t)$ in the action. In addition, the tensor perturbation  $\gamma_{ij} $ is always accompanied 
by $a^{-2}(t)$ since  $\gamma$ comes  from the inverse spatial metric $h^{ij}$, which has $a^{-2}(t)$. 
This explains the power of $a(t)$ in   (\ref{hamihami1}),  (\ref{hamihami2}) and  (\ref{hamihami3}). 
Also note that we are allowed to impose $\parti\partj$, which operated on $\sigma$ in (\ref{cubicgzsig}), on $\zetadot$ because in the soft-graviton limit, $\sigma$ and $\zetadot$ have the same momentum in the momentum space.
Finally, notice that there is a factor `$2$' in the last in (\ref{hamihami2}) because there are two diagrams originally as Figure \ref{effectivefigure} [II] shows. \vspace{3mm}
%%%%%%%%%%%%%%%%%%%%%%%%%%%%%%%%%%%%%%%
%%%%%%%%%%%%%%%%%%%%%%%%%%%%%%%%%%%%%%%

Since the simplified diagrams (those on the right hand side  in Figure \ref{effectivefigure} ) have only one vertex, the computation 
is much easier than the original ones (left ones in Figure \ref{effectivefigure}). 
Using the in-in formalism  (\ref{eq:expectationvalueO}), 
% (\ref{ininformula}), 
the correlation function of $\mathcal{O}(t,\vx)$ ($=\zeta\zeta$, $\gamma\zeta\zeta$ and $\gamma\gamma\zeta\zeta$)  in this case is given in the first order  by
\begin{equation}
%\begin{split}
\langle 0(t) \vert \mathcal{O}(t,\vx) \vert 0(t) \rangle \Big \vert_{{\rm 1st. \: order}} 
=2\,\textrm{Im}\left[\int_{-\infty}^{t}dt'\ \langle 0 \vert \mathcal{O}(t,\vx)H_{I}(t') 
\vert 0 \rangle \right].
%\end{split}
\label{inineffective}
\end{equation}
Putting  the interaction Hamiltonian (\ref{hamihami1}),(\ref{hamihami2}) and (\ref{hamihami3}) into (\ref{inineffective}) , and also using the two point functions discussed previously, 
we can reach the following simplified formulae:\vspace{3mm}
%%%%%%%%%%%%%%%%%%%%%%%%%%%%%%%%%%%%%%%%%%%%%%%%%%%%%%%%%%%%%%%%%%%%%%%%%%%%%%%%%%%%%%
%\leftline{\underline{{\bf [I] Computation of} $\langle \zeta \zeta \rangle $}}
\begin{eqnarray}
\langle\zeta(\vk_1)\zeta(\vk_2)\rangle \Big\vert_{{\rm Fig. \: \ref{effectivefigure} \: [I]}} 
& \Rightarrow&
\quad(2\pi)^3 \delta^3 (\vk_1+\vk_2)
\nonumber \\
& &\qquad\times f(m)\left(-2\frac{c_1}{H}\right)^2\left(\biarii\right)\left(\biari\right)
\times\,2!
\nonumber \\
& &\qquad\qquad\times2\,\textrm{Im}\left[\int_{-\infty}^{0
}\frac{d\eta}{\eta^4 H^4}\eta^2 e^{ik_1\eta}\eta^2 e^{ik_2\eta}\right]
\nonumber \\
&=&
\quad(2\pi)^3 \delta^3 (\vk_1+\vk_2)\,P_{\zeta}(k_2)\,\frac{c_1^2}{\epsilon H^4}\tilde{F}_1(m),
\label{f1til}
\end{eqnarray}
%\begin{equation}
%\label{f1tildif}
%\left(\tilde{F}_1 (m)\equiv-2H^2 f(m)\right)
%\end{equation}
%
%\leftline{\underline{{\bf [I\hspace{-.1em}I] Computation of} $\langle \gamma \zeta \zeta \rangle $ {\bf in the soft-graviton limit} $\vq\to0$}}
\begin{eqnarray}
\langle\gamma^{s}(\vq)\zeta(\vk_1)\zeta(\vk_2)\rangle
\Big\vert_{{\rm Fig. \: \ref{effectivefigure} \: [II]}}
&\Rightarrow &
\quad(2\pi)^3\delta^3 (\vq+\vk_1+\vk_2)\ \epsilon_{ij}^{s}(\vq)(-i)^2 (k_2)_{i}(k_2)_{j}
\nonumber \\
& &\qquad\times f(m)\left(-4\frac{c_1 c_4}{H^2}\right)\binashiq\left(\biarii\right)\left(\biari\right)\times2\,\times\,2!
\nonumber \\
& &\qquad\qquad
\times2\,\textrm{Im}\left[\int_{-\infty}^{0}\frac{d\eta}{\eta^2 H^2}\eta^2 e^{ik_1\eta}\eta^2 e^{ik_2\eta}\right]
\nonumber \\
&=&
\quad(2\pi)^3 
\delta^3 (\vq+\vk_1+\vk_2)\ \epsilon_{ij}^{s}(\vq)\frac{(k_2)_{i}(k_2)_{j}}{(k_2)^2}
\nonumber \\
& &\times P_{\gamma}(q)P_{\zeta}(k_2) \frac{c_1 c_4}{\epsilon H^2}\,\tilde{F}_2(m),
\label{f2til}
\end{eqnarray}
%\begin{equation}\label{f2tildif}
%\left(\tilde{F}_2 (m)\equiv2H^2 f(m)\right)
%\end{equation}
%
%\leftline{\underline{{\bf [I\hspace{-.1em}I\hspace{-.1em}I] Computation of} $\langle  \gamma \gamma \zeta \zeta$ {\bf in the soft-graviton limit} $\vq_1,\vq_2\to0$}}
\begin{eqnarray}
\langle\gamma^{s_1}(\vq_1)\gamma^{s_2}(\vq_2)\zeta(\vk_1)\zeta(\vk_2)\rangle
\Big\vert_{{\rm Fig. \: \ref{effectivefigure} \: [III]}}
& \Rightarrow &
\quad (2\pi)^3\delta^3 (\vq_1+\vq_2+\vk_1+\vk_2)
\nonumber \\
& &\times\epsilon_{ij}^{s_1}(\vq_1)(-i)^2 (k_1)_{i}(k_1)_{j}\ \epsilon_{\alpha\beta}^{s_2}(\vq_2)(-i)^2 (k_2)_{\alpha}(k_2)_{\beta}
\nonumber \\
& &\times f(m)\left(2\frac{c_4}{H}\right)^2\frac{H^2}{(q_1)^3}\frac{H^2}{(q_2)^3}
\left(\biarii\right)\left(\biari\right)
\nonumber \\
& &\quad \times\,2!\times2!
%\nonumber \\
%& &\qquad
\times2\,\textrm{Im}\left[\int_{-\infty}^0 d\eta\ \eta^2 e^{ik_1\eta}\eta^2 e^{ik_2\eta}\right]
\nonumber \\
&=&
\quad (2\pi)^3\delta^3 (\vq_1+\vq_2+\vk_1+\vk_2)
\nonumber \\
& &\quad\times\epsilon_{ij}^{s_1}(\vq_1)\frac{(k_2)_{i}(k_2)_{j}}{(k_2)^2}\epsilon_{\alpha\beta}^{s_2}(\vq_2)\frac{(k_2)_{\alpha}(k_2)_{\beta}}{(k_2)^2}
\nonumber \\
& &\qquad\times P_{\gamma}(q_1)P_{\gamma}(q_2)P_{\zeta}(k_2)\frac{c_4^2}{\epsilon}\,\tilde{F}_3(m), 
\label{f3til}
\end{eqnarray}
%\begin{equation}\label{f3tildif}
%\left(\tilde{F}_3 (m)\equiv-6H^2 f(m)\right)
%\end{equation}
where
\begin{equation}
{\tilde F}_{1}(m)=-2H^{2}f(m), \qquad {\tilde F}_{2}(m)=2H^{2}f(m), \qquad {\tilde F}_{3}(m)=-6H^{2}f(m)\:. 
\label{eq:f123til}
\end{equation}
%%%%%%%%%%%%%%%%%%%%%%%%%%%%%%%%%%%%%%%%%%%%%%%%%%%%%%%%%%%%%%%%%%%%%%%%%%%%%%%%%%%%%%%%%%%%%%%%%%
Note that we have used 
%the integration by parts and 
the $i\epsilon$ prescription when we compute the integrals above (see Appendix \ref{app:C} ).

Comparing (\ref{f1til}), (\ref{f2til}) and (\ref{f3til}) with (\ref{f1ori}), (\ref{f2ori}) and (\ref{f3ori}), it follows that $\tilde{F}_1(m)$, $\tilde{F}_2(m)$ and $\tilde{F}_3(m)$ correspond to $F_1(m)$, $F_2(m)$ and $F_3(m)$,  
respectively in the $m\to\infty$ limit. On looking at  (\ref{eq:f123til}), we can immediately see  simple relations 
\begin{eqnarray}
\tilde{G}_{12}(m)&\equiv&\frac{\tilde{F}_1(m)}{\tilde{F}_2(m)}\quad=\quad-1,\label{G12rela}\\
\tilde{G}_{23}(m)&\equiv&\frac{\tilde{F}_2(m)}{\tilde{F}_3(m)}\quad=\quad-\frac{1}{3}.\label{G23rela}
\end{eqnarray}
These results are consistent  with the numerical analysis of the original diagrams shown in 
Table \ref{numericalresult} 
and agree with (\ref{eq:asymptoticvalues}).
The relations in  (\ref{eq:asymptoticvalues}) 
 may be useful in order to search for particles whose masses are much 
greater  than $H\sim10^{13} - 10^{14}$ GeV; although each function $F_1(m)$, $F_2(m)$ and $F_3(m)$ approaches zero as Figures \ref{fonem}, \ref{ftwom} and \ref{fthreem} show, we may get a hint of unknown particles by 
measuring the ratios of correlation functions.

%%%%%%%%%%%%%%%%%%%%%%%%%%%%%%%%%%%%%%%%%%%%%%%%%
\section{The generalization to the correlations with $N$-gravitons }
\label{section5}

We have seen in the previous section that the contributions  due to the new coupling 
(\ref{cubicgzsig}) 
to the correlation functions 
\begin{eqnarray}
\langle \gamma^{s} \zeta \zeta \rangle \sim c_{1\: }c_{4}\: F_{2}(m), 
\qquad
\langle \gamma^{s_{1} } \gamma ^{s_{2}} \zeta \zeta \rangle \sim (c_{4})^{2}\: F_{3}(m)
\end{eqnarray}
are closely connected with each other for large $m$ by virtue of the relation 
$G_{23}(m)=F_{2}(m)/F_{3}(m) \to -1/3$ as $m \to \infty$. In the present section we would like to 
point out that we can generalize this 
curious connection to that of the following  correlation functions 
\begin{eqnarray}
\langle \gamma^{s_{1}}  \cdots \gamma ^{s_{N}} \zeta \zeta \rangle , 
\qquad
%\langle \gamma^{s_{1}}  \cdots \gamma ^{s_{N+1}} \zeta \zeta \rangle ,  \qquad
\langle \gamma^{s_{1} } \cdots \cdots \gamma ^{s_{2N}} \zeta \zeta \rangle . 
\end{eqnarray}

\begin{figure}[H]
  \centering
  \includegraphics[width=9cm]{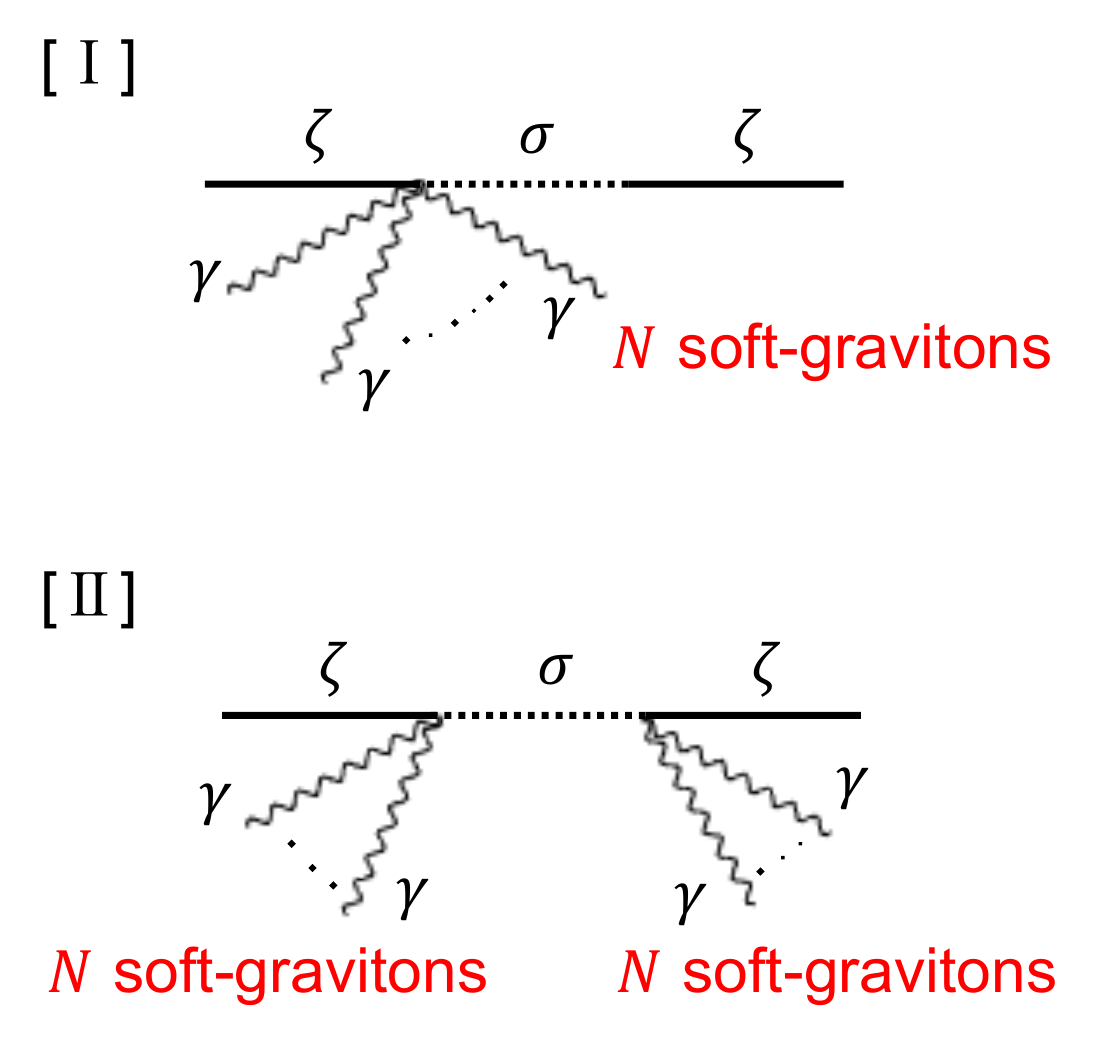}
  \caption{[I] `$(N,0)$ soft-gravitons' diagram\  and [I\hspace{-.1em}I]  `$(N,N)$ soft-gravitons' diagram}
  \label{General}
\end{figure}

In order to compute the correlation functions involving  an arbitrary number $N$ soft-graviton vertex, 
let us consider the action below containing more $g^{\mu \nu}$'s  than in  (\ref{eftaction3}):
\begin{equation}
\begin{split}
S_{I}=\int d^4 x\sqrt{-g}\,\delta g^{00}\biggl[C_0 \sigma&+C_1 g^{\mu\nu}\partial_{\mu}\partial_{\nu}\sigma 
%\\&
+C_2g^{\mu\nu}g^{\rho\tau}\partial_{\mu}\partial_{\nu}\partial_{\rho}\partial_{\tau}\sigma+\cdots\biggl].
\end{split}
\end{equation}
Note that $C_0$ and $C_1$ correspond to $c_1$ and $c_4$ in 
 (\ref{eftaction3}),  respectively. If we consider just one graviton for each $g^{\mu\nu}$, it follows that the term 
involving  $C_{N}$ produces a coupling of $\zeta$, $\sigma$ and $N$ soft-gravitons. Therefore, using $\delta g^{00}=2\zetadot/H$ and $h^{ij}=-a^{-2}(t)\gaij$ to the first order, the interaction Hamiltonian $H_{I}(t)=-L_{I}(t)$ becomes
\begin{equation}
\begin{split}
H_{I}(t)=&-2\frac{C_0}{H}a^3(t)\int d^3 x\,\zetadot\sigma\\
&+2\frac{C_1}{H}a(t)\int d^3 x\,\zetadot\gaij\parti\partj\sigma\\
&-2\frac{C_2}{H}a^{-1}(t)\int d^3 x\,\zetadot\gaij\gamma_{\alpha\beta}\parti\partj\partial_{\alpha}\partial_{\beta}\sigma\\
&+\cdots\\
&+(-1)^{N+1}2\frac{C_{N}}{H}a^{3-2N}(t)\int d^3 x\,\zetadot\gaij\cdots\gamma_{\alpha\beta}\parti\partj\cdots\partial_{\alpha}\partial_{\beta}\sigma\\
&+\cdots.
\end{split}
\label{eq:interactionofNgravitons}
\end{equation}
Using the interaction  Hamiltonian associated with $N$-gravitons 
 once and twice,   we will compute two types of the correlation functions 
shown  in Figure \ref{General}. 
The diagrams [I] and [I\hspace{-.1em}I] in Figure \ref{General} will be  called 
`$(N, 0)$ and $(N, N)$ soft-graviton diagrams',   respectively.
They are the generalization of [I\hspace{-.1em}I] and [I\hspace{-.1em}I\hspace{-.1em}I] in 
Figure \ref{diagrams}. 
%%%%%%%%%%%%%%%%%%%%%%%%%%%%%%%%%%%%%%%%
%%%%%%%%%%%%%%%%%%%%%%%%%%%%%%%%%%%%%%%%%%

%%%%%%%%%%%%%%%%%%%%%%%%%%%%%%%%%%%%%%%%%%%%%%%%%%%%%
\subsection{General Considerations}
%\subsection{Computations and results}
\leftline{\textbf{[I] Computation of the `$(N,0)$ soft-gravitons' diagram}}\vspace{2mm}
The strategy for the computation of the `$(N, 0)$ soft-graviton diagram' is the same as in 
 (\ref{gzzmiddle}), and the calculation goes as follows:
\begin{equation}
\begin{split}
&\langle\gamma^{s_1}(\vq_1)\cdots\gamma^{s_{N}}(\vq_{N})\zeta(\vk_1)\zeta(\vk_2)\rangle
\Big \vert _{{\rm Fig}. \: \ref{General}\: [ {\rm I} ]}
\\
&=(2\pi)^3\delta^3 (\vq_1+\cdots +\vq_{N}+\vk_1+\vk_2)\\
&\qquad\times\epsilon_{ij}^{s_1}(\vq_1)\cdots\epsilon_{\alpha\beta}^{s_{N}}(\vq_{N})(-i)^2\cdots(-i)^2 (k_2)_{i}(k_2)_{j}\cdots(k_2)_{\alpha}(k_2)_{\beta}\\
&\qquad\times(-1)^{N}4\frac{C_0 C_{N}}{H^2}\frac{H^2}{q_1^3}\cdots\frac{H^2}{q_{N}^3}\left(\biarii\right)\left(\biari\right)\\
&\qquad\times\sigmatip\, \times\,2!\times N!   \\
&\times\Biggl[2\textrm{Re}\left[\int_{-\infty}^{0}\frac{d\eta_1}{\eta_1^4H^4}\int_{-\infty}^{0}\frac{d\eta_2}{(\eta_2 H)^{4-2N}}\eta_1^2e^{-ik_1\eta_1}\eta_2^2e^{ik_2\eta_2}(\eta_1\eta_2)^{\frac{3}{2}}\hankel(-k_2\eta_1)\hankelc(-k_2\eta_2)\right]\\
&\quad-2\textrm{Re}\left[\int_{-\infty}^{0}\frac{d\eta_1}{\eta_1^4H^4}\int_{-\infty}^{\eta_1}\frac{d\eta_2}{(\eta_2 H)^{4-2N}}\eta_1^2e^{ik_1\eta_1}\eta_2^2e^{ik_2\eta_2}(\eta_1\eta_2)^{\frac{3}{2}}\hankel(-k_2\eta_1)\hankelc(-k_2\eta_2)\right]\\
&\quad-2\textrm{Re}\left[\int_{-\infty}^{0}\frac{d\eta_1}{(\eta_1 H)^{4-2N}}\int_{-\infty}^{\eta_1}\frac{d\eta_2}{\eta_2^4H^4}\eta_1^2e^{ik_1\eta_1}\eta_2^2e^{ik_2\eta_2}(\eta_1\eta_2)^{\frac{3}{2}}\hankel(-k_2\eta_1)\hankelc(-k_2\eta_2)\right]\Biggl].
\end{split}\label{geneonemid}
\end{equation}
The  factor $2! \times N !$ in the fifth line 
is a combinatorial factor for the diagram [I] in Figure \ref{General}.\vspace{2mm}
We then set  $x\equiv-k_2 \eta_1$ and $y\equiv -k_2 \eta_2$ to find 
\begin{equation}
\begin{split}
&\langle\gamma^{s_1}(\vq_1)\cdots\gamma^{s_{N}}(\vq_{N})\zeta(\vk_1)\zeta(\vk_2)\rangle
\Big \vert _{{\rm Fig}. \: \ref{General}\: [ {\rm I} ]}
\\
&=(2\pi)^3\delta^3 (\vq_1+\cdots +\vq_{N}+\vk_1+\vk_2)\\
&\qquad\times\epsilon_{ij}^{s_1}(\vq_1)\cdots\epsilon_{\alpha\beta}^{s_{N}}(\vq_{N})\frac{(k_2)_{i}(k_2)_{j}}{(k_2)^2}\cdots\frac{(k_2)_{\alpha}(k_2)_{\beta}}{(k_2)^2}\\
&\qquad\times P_{\gamma}(q_1)\cdots P_{\gamma}(q_{N})\ P_{\zeta}(k_2)\frac{\pi C_0 C_{N}}{\epsilon H^{4-2N}}N!\,e^{-\pi\textrm{Im}(\nu)}\\
&\qquad\times\Biggl[\textrm{Re}\left[\int_{0}^{\infty}dx\,x^{-\frac{1}{2}}e^{ix}\hankel(x)\int_{0}^{\infty}dy\,y^{2N-\frac{1}{2}}e^{-iy}\hankelc(y)\right]\\
&\qquad\quad-\textrm{Re}\left[\int_{0}^{\infty}dx\,x^{-\frac{1}{2}}e^{-ix}\hankel(x)\int_{x}^{\infty}dy\,y^{2N-\frac{1}{2}}e^{-iy}\hankelc(y)\right]\\
&\qquad\quad-\textrm{Re}\left[\int_{0}^{\infty}dx\,x^{2N-\frac{1}{2}}e^{-ix}\hankel(x)\int_{x}^{\infty}dy\,y^{-\frac{1}{2}}e^{-iy}\hankelc(y)\right]\Biggl].
\end{split}\label{geneoneres}
\end{equation}
We can compute these integrals just by using the formulae (\ref{resulteasy}) and (\ref{resultdiff}) in 
Appendices \ref{app:A} and \ref{app:B} and we obtain finally 
\begin{equation}
\begin{split}
\langle\gamma^{s_1}(\vq_1)&\cdots\gamma^{s_{N}}(\vq_{N})\zeta(\vk_1)\zeta(\vk_2)\rangle
\Big \vert _{{\rm Fig}. \: \ref{General}\: [ {\rm I} ]}
\\
&=(2\pi)^3\delta^3 (\vq_1+\cdots +\vq_{N}+\vk_1+\vk_2)\\
&\qquad\times\epsilon_{ij}^{s_1}(\vq_1)\cdots\epsilon_{\alpha\beta}^{s_{N}}(\vq_{N})\frac{(k_2)_{i}(k_2)_{j}}{(k_2)^2}\cdots\frac{(k_2)_{\alpha}(k_2)_{\beta}}{(k_2)^2}\\
&\qquad\times P_{\gamma}(q_1)\cdots P_{\gamma}(q_{N})\ P_{\zeta}(k_2)\frac{C_0 C_{N}}{\epsilon H^{4-2N}}\,R_{N}(m),
\end{split}\label{geneoneresres}
\end{equation}
where we have introduced  a function 
\begin{equation}
\begin{split}
&R_{N}(m)\equiv\\
&\,2^{1-2N}(-1)^{N}N!\\
&\times\Biggl[\frac{\pi^2}{\cosh^2(\pi\mu)}\frac{1}{(2N)!}\left\{\left(2N-\frac{1}{2}\right)^2+\mu^2\right\}\cdots\left\{\left(\frac{3}{2}\right)^2+\mu^2\right\}\left\{\left(\frac{1}{2}\right)^2+\mu^2\right\}\\
&\quad-\frac{1}{\sinh(\pi\mu)}\,\textrm{Re}\Biggl[\,\sum_{n=0}^{\infty}(-1)^{n}(n+1)(n+2)\cdots(n+2N)\\
&\qquad\times\Biggl\{e^{\pi\mu}\frac{(1+n+2i\mu)(2+n+2i\mu)\cdots(2N+n+2i\mu)\times(N+\frac{1}{2}+n+i\mu)}{(\frac{1}{2}+n+i\mu)^2(\frac{3}{2}+n+i\mu)\cdots(2N-\frac{1}{2}+n+i\mu)(2N+\frac{1}{2}+n+i\mu)^2}\\
&\qquad -e^{-\pi\mu}\frac{(1+n-2i\mu)(2+n-2i\mu)\cdots(2N+n-2i\mu)\times(N+\frac{1}{2}+n-i\mu)}{(\frac{1}{2}+n-i\mu)^2(\frac{3}{2}+n-i\mu)\cdots(2N-\frac{1}{2}+n-i\mu)(2N+\frac{1}{2}+n-i\mu)^2}\Biggl\}\Biggl]\Biggl].
\end{split}\label{INmuzui}
\end{equation}
This result is consistent with (\ref{f2muzui}) for the  $N=1$  case. 
Actually we can confirm that the function $R_{1}(m)$ coincides with $F_{2}(m)$ of (\ref{f2muzui}).

\vskip0.3cm
%%%%%%%%%%%%%%%%%%%%%%%%%%%%%%%%%%%%%%%%%%%%%%%%%%%%%%%%%%%%%%%%
\leftline{\textbf{[I\hspace{-.1em}I] Computation of the `$(N,N)$ soft-gravitons' diagram}}\vspace{2mm}

\noindent
This computation of the `$(N,N)$ soft-gravitons' diagram
goes along the  same line as for (\ref{ggzzmiddle}) and we get
\begin{equation}
\begin{split}
&\langle\gamma^{s_1}(\vq_1)\cdots\gamma^{s_{2N}}(\vq_{2N})\zeta(\vk_1)\zeta(\vk_2)\rangle
\Big \vert _{{\rm Fig}. \: \ref{General}\: [ {\rm II } ]}
\\
&=(2\pi)^3\delta^3 (\vq_1+\cdots +\vq_{2N}+\vk_1+\vk_2)\\
&\qquad\times\epsilon_{ij}^{s_1}(\vq_1)\cdots\epsilon_{\alpha\beta}^{s_{2N}}(\vq_{2N})(-i)^2\cdots(-i)^2 (k_2)_{i}(k_2)_{j}\cdots(k_2)_{\alpha}(k_2)_{\beta}\\
&\qquad\times4\frac{C_{N}^2}{H^2}\frac{H^2}{q_1^3}\cdots\frac{H^2}{q_{2N}^3}\left(\biarii\right)\left(\biari\right)\\
&\qquad\times\sigmatip\,
\times\,2!\times (2N)!
\\
&\times\Biggl[\int_{-\infty}^{0}\frac{d\eta_1}{(\eta_1 H)^{4-2N}}\int_{-\infty}^{0}\frac{d\eta_2}{(\eta_2 H)^{4-2N}}\eta_1^2e^{-ik_1\eta_1}\eta_2^2e^{ik_2\eta_2}(\eta_1\eta_2)^{\frac{3}{2}}\hankel(-k_2\eta_1)\hankelc(-k_2\eta_2)\\
&-2\textrm{Re}\left[\int_{-\infty}^{0}\frac{d\eta_1}{(\eta_1 H)^{4-2N}}\int_{-\infty}^{\eta_1}\frac{d\eta_2}{(\eta_2 H)^{4-2N}}\eta_1^2e^{ik_1\eta_1}\eta_2^2e^{ik_2\eta_2}(\eta_1\eta_2)^{\frac{3}{2}}\hankel(-k_2\eta_1)\hankelc(-k_2\eta_2)\right]\Biggl].
\end{split}\label{genetwomid}
\end{equation}
As before,  the  factor $2! \times (2N)!$ in the fifth line 
is a combinatorial factor for the diagram [I\hspace{-.1em}I] in Figure \ref{General}.\vspace{2mm}
By setting $x\equiv-k_2 \eta_1$ and $y\equiv -k_2 \eta_2$, we can put (\ref{genetwomid}) into the following form, 
\begin{equation}
\begin{split}
&\langle\gamma^{s_1}(\vq_1)\cdots\gamma^{s_{2N}}(\vq_{2N})\zeta(\vk_1)\zeta(\vk_2)\rangle
\Big \vert _{{\rm Fig}. \: \ref{General}\: [ {\rm II } ]}
\\
&=(2\pi)^3\delta^3 (\vq_1+\cdots +\vq_{2N}+\vk_1+\vk_2)\\
&\qquad\times\epsilon_{ij}^{s_1}(\vq_1)\cdots\epsilon_{\alpha\beta}^{s_{2N}}(\vq_{2N})\frac{(k_2)_{i}(k_2)_{j}}{(k_2)^2}\cdots\frac{(k_2)_{\alpha}(k_2)_{\beta}}{(k_2)^2}\\
&\qquad\times P_{\gamma}(q_1)\cdots P_{\gamma}(q_{2N})\ P_{\zeta}(k_2)\frac{\pi C_{N}^2}{\epsilon H^{4(1-N)}}(2N)!\,e^{-\pi\textrm{Im}(\nu)}\\
&\qquad\times\Biggl[\frac{1}{2}\left|\int_{0}^{\infty}dx\,x^{2N-\frac{1}{2}}e^{ix}\hankel(x)\right|^2\\
&\qquad\quad-\textrm{Re}\left[\int_{0}^{\infty}dx\,x^{2N-\frac{1}{2}}e^{-ix}\hankel(x)\int_{x}^{\infty}dy\,y^{2N-\frac{1}{2}}e^{-iy}\hankelc(y)\right]\Biggl].
\end{split}\label{genetwores}
\end{equation}
Finally we make use of the integration formulae (\ref{resulteasy}) and (\ref{resultdiff}) in Appendices  \ref{app:A} and 
\ref{app:B}
 and this correlation function turns out to be 
\begin{equation}
\begin{split}
\langle\gamma^{s_1}(\vq_1)&\cdots\gamma^{s_{2N}}(\vq_{2N})\zeta(\vk_1)\zeta(\vk_2)\rangle
\Big \vert _{{\rm Fig}. \: \ref{General}\: [ {\rm II } ]}
\\
&=(2\pi)^3\delta^3 (\vq_1+\cdots +\vq_{2N}+\vk_1+\vk_2)\\
&\qquad\times\epsilon_{ij}^{s_1}(\vq_1)\cdots\epsilon_{\alpha\beta}^{s_{2N}}(\vq_{2N})\frac{(k_2)_{i}(k_2)_{j}}{(k_2)^2}\cdots\frac{(k_2)_{\alpha}(k_2)_{\beta}}{(k_2)^2}\\
&\qquad\times P_{\gamma}(q_1)\cdots P_{\gamma}(q_{2N})\ P_{\zeta}(k_2)\frac{C_{N}^2}{\epsilon H^{4(1-N)}}\,S_{N}(m),
\end{split}\label{genetworesres}
\end{equation}
where we have defined a function 
\begin{equation}
\begin{split}
&S_{N}(m)\equiv\\
&2^{-4N}\Biggl[\frac{\pi^2}{\cosh^2(\pi\mu)}\frac{1}{(2N)!}\left\{\left(2N-\frac{1}{2}\right)^2+\mu^2\right\}^2\cdots\left\{\left(\frac{3}{2}\right)^2+\mu^2\right\}^2\left\{\left(\frac{1}{2}\right)^2+\mu^2\right\}^2\\
&\quad-\frac{(2N)!}{\sinh(\pi\mu)}\,\textrm{Re}\Biggl[\,\sum_{n=0}^{\infty}(-1)^{n}(n+1)(n+2)\cdots(n+4N)\\
&\qquad\times\Biggl\{e^{\pi\mu}\frac{(1+n+2i\mu)(2+n+2i\mu)\cdots(4N+n+2i\mu)}{(\frac{1}{2}+n+i\mu)(\frac{3}{2}+n+i\mu)\cdots(2N+\frac{1}{2}+n+i\mu)^2\cdots(4N+\frac{1}{2}+n+i\mu)}\\
&\qquad -e^{-\pi\mu}\frac{(1+n-2i\mu)(2+n-2i\mu)\cdots(4N+n-2i\mu)}{(\frac{1}{2}+n-i\mu)(\frac{3}{2}+n-i\mu)\cdots(2N+\frac{1}{2}+n-i\mu)^2\cdots(4N+\frac{1}{2}+n-i\mu)}\Biggl\}\Biggl]\Biggl].
\end{split}\label{JNmuzui}
\end{equation}
This result is consistent with (\ref{f3muzui}) for the  $N=1$ case.
 In fact  we can confirm that the function $S_{1}(m)$ coincides with $F_{3}(m)$ of (\ref{f3muzui}).

%%%%%%%%%%%%%%%%%%%%%%%%%%%%%%%%%%%%%%%%%%%%%%%%%%%%%%
\subsection{Evaluation of the functions $R_{N}(m)$ and $S_{N}(m)$}

\begin{figure}[H]
  \centering
  \includegraphics[width=11cm]{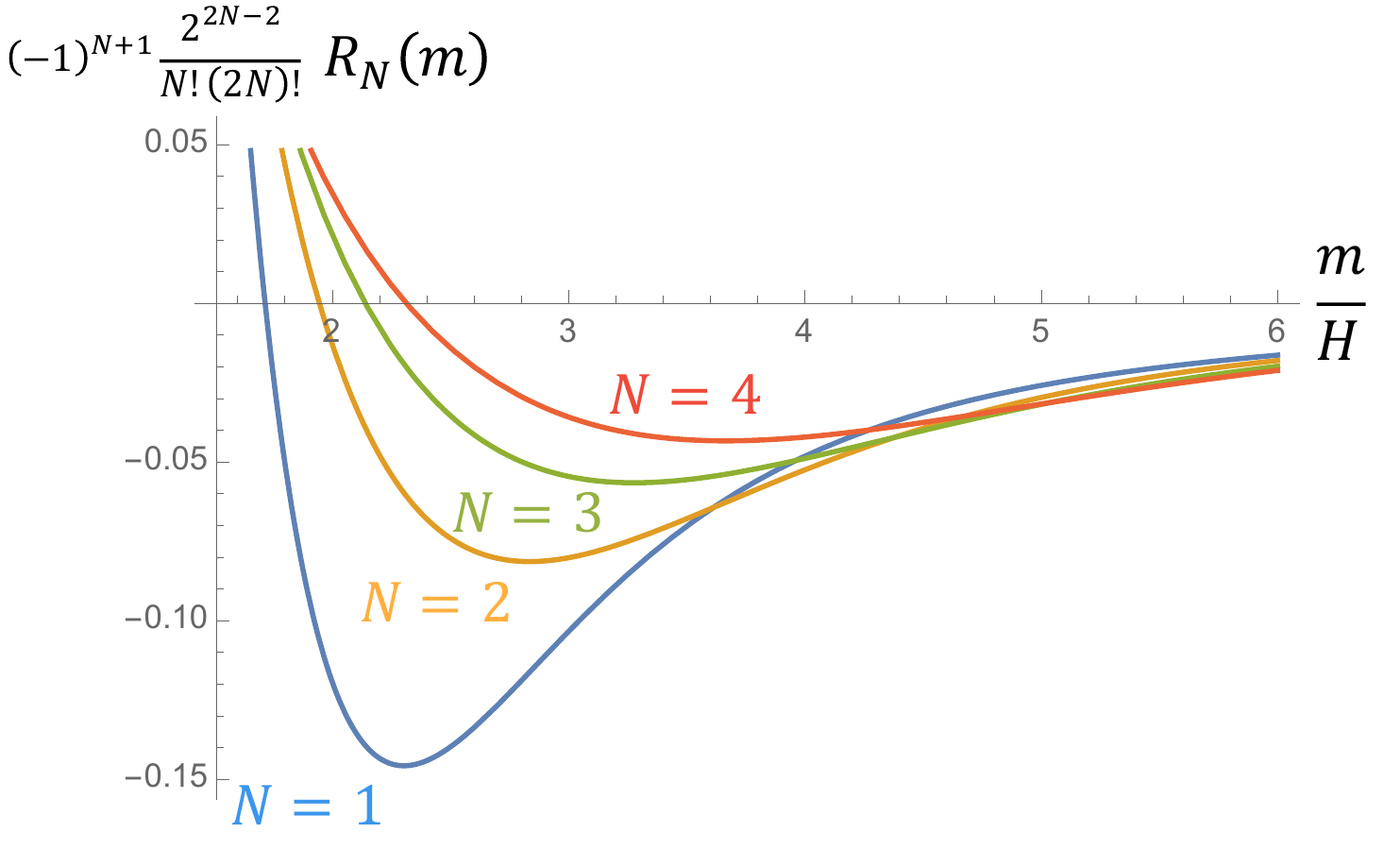}
  \caption{The plots for $(-1)^{N+1}\frac{2^{2N-2}}{N!(2N)!}R_{N}(m)$ when $N=1,2,3$ and $4$ in $m\geq1.5H$. The factor $(-1)^{N+1}\frac{2^{2N-2}}{N!(2N)!}$ is introduced for the functions to converge to the same value when $m\to\infty$; we will discuss this in the  subsection
\ref{subsection:largemassbehavior}, 
and this factor is shown in (\ref{Itildef}).}
  \label{INfour}
\end{figure}

To have an insight into the sensitivity of the correlation 
$\langle \gamma ^{s_{1}} \cdots \gamma ^{s_{N}} \zeta \zeta \rangle$
to the mass $m$, let us consider   the functions  $R_{N}(m)$ for $N=1,2,3$ and $4$,  whose  
 $m$-dependence are illustrated 
  in Figure \ref{INfour}.
As we can see easily in Figure \ref{INfour} , the peak of the correlation function is shifted to 
larger values of the mass $m$ of $\sigma$ as the number of the soft-gravitons increases. \vspace{2mm}

%%%%%%%%%%%%%%%%%%%%%%%%%%%%%%%%%%
%%%%%%%%%%%%%%%%%%%%%%%%%%%%%%%%%%%%%

%%%%%%%%%%%%%%%%%%%%%%%%%%%
%\begin{figure}[H]
%  \centering
%  \includegraphics[width=9cm]{INdiagrams.pdf}
%  \caption{The diagrams for $R_{N}(m)$ when $N=1,2,3$ and $4$}
%  \label{INdiagrams}
%\end{figure}
%%%%%%%%%%%%%%%%%%%%%%%%%%%%%%%%%%%

\begin{figure}[H]
  \centering
  \includegraphics[width=11cm]{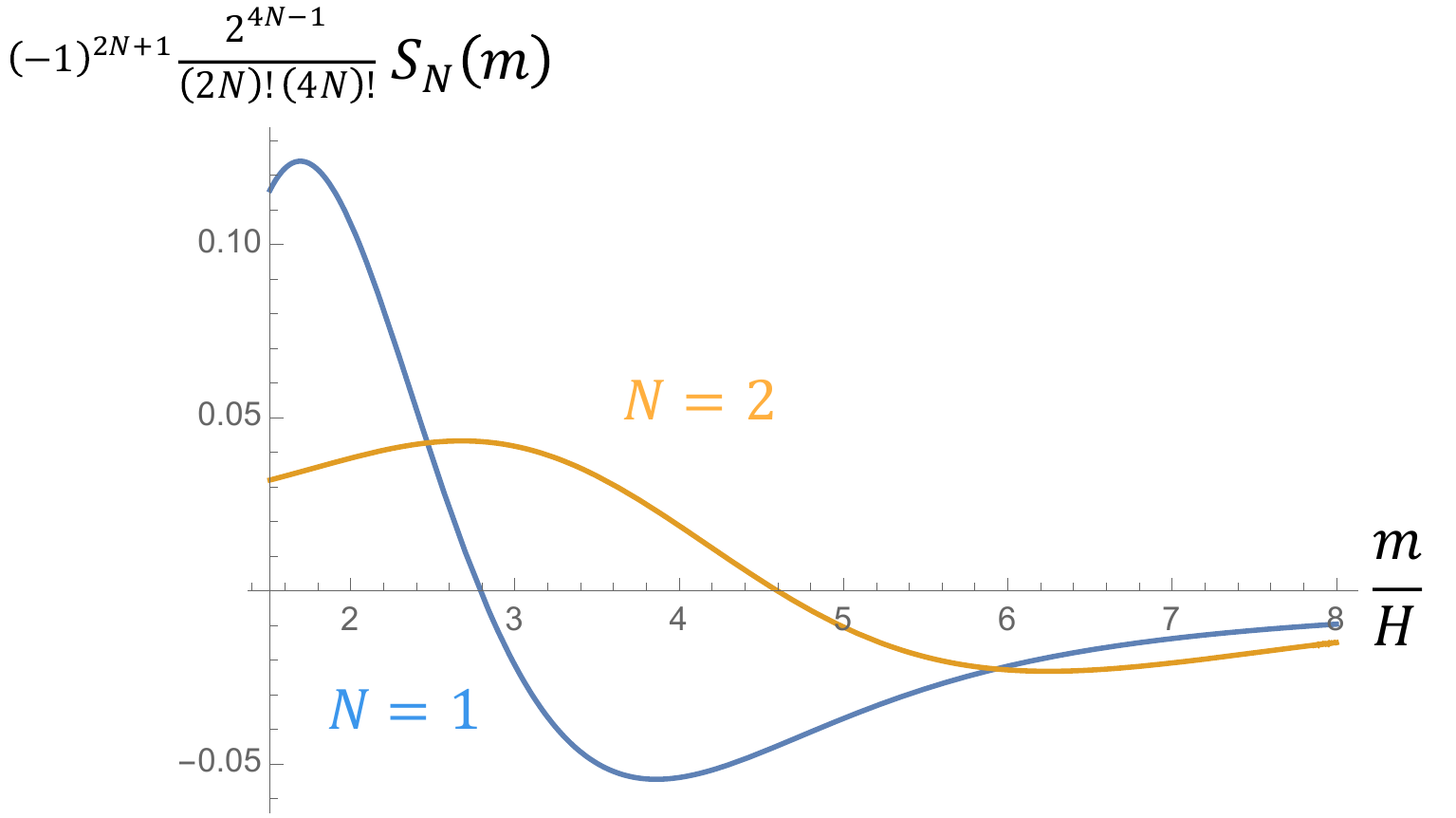}
  \caption{The plots for $(-1)^{2N+1}\frac{2^{4N-1}}{(2N)!(4N)!}S_{N}(m)$ when $N=1$ and $2$ in $m\geq1.5H$. The factor $(-1)^{2N+1}\frac{2^{4N-1}}{(2N)!(4N)!}$ is introduced for the same reason as Figure \ref{INfour}, and it is shown in (\ref{Jtildef}).}
  \label{JNtwo}
\end{figure}

Similarly we also plot the behavior of the function 
of  $S_{N}(m)$ for  $N=1$ and $2$ in Figure \ref{JNtwo}. 
Although we were able to study only two cases $N=1 $ and $N=2$, 
because of the limited computational power, it is likely that  
 the positions of the peaks are shifted to larger values of the mass $m$ of $\sigma$ as 
 $N$ becomes large. 
We may safely say that correlation functions involving larger number of gravitons would be useful 
for probing larger values of $m$. 
\vspace{2mm}
%%%%%%%%%%%%%%%%%%%%%%%%%%%%%%%%%%%%%%%
%%%%%%%%%%%%%%%%%%%%%%%%%%%%%%%%%%%%%%%%%%
%%%%%%%%%%%%%%%%%%%%%%%%%%%%%%%%%%%%%%
%\begin{figure}[H]
%  \centering
%  \includegraphics[width=9cm]{JNdiagrams.pdf}
%  \caption{The diagrams for $S_{N}(m)$ when $N=1$ and $2$}
%  \label{JNdiagrams}
%\end{figure}
%%%%%%%%%%%%%%%%%%%%%%%%%%%%%%%%%%%%%%%%

We are able to see from Figures \ref{INfour}  and \ref{JNtwo} that the functions $R_{N}(m)$ and $S_{N}(m)$ 
both go down quickly for large $m$, and that 
the speed of the decrease seems to be shared among them.
As one may  surmise from (\ref{INmuzui}) and (\ref{JNmuzui}), 
a numerical check of the speed of their decrease is not easy to do for large $N$.  
Therefore, here we examine only the ratios of $R_{1}(m)$, $R_2(m)$, $R_3(m)$ and $S_1(m)$ 
for  $m \gg H$ and some of the results are listed in Tables \ref{numericalresult2} and \ref{numericalresult3}. 
We can safely conclude from the numbers listed in Table \ref{numericalresult2} that
the limiting behavior is  $R_2(m)/R_1(m)\to-6$ and $R_2(m)/S_1(m)\to2$ as $m/H \to\infty$.
In the next subsection we will develop a method to explain these limits, 
thereby the limiting value suggested  in 
 Table \ref{numericalresult3} will  also be explained.

%%%%%%%%%%%%%%%%%%%%%%%%%%
\begin{table}[H]
  \centering
  \caption{The numerical analysis of $R_2(m)/R_1(m)$ and $R_2(m)/S_1(m)$}
  \begin{tabular}{|c||c|c|c|c|c|c|}\hline
  $\displaystyle m/H$&$\displaystyle 5.0$&$\displaystyle 10.0$&$\displaystyle 15.0$&$\displaystyle 20.0$&$\displaystyle 25.0$&$\displaystyle 30.0$\\ \hline\hline
  $\displaystyle R_2(m)/R_1(m)$&$\displaystyle -6.88167$&$\displaystyle -6.14334$&$\displaystyle -6.05725$&$\displaystyle -6.03114$&$\displaystyle -6.02061$&$\displaystyle -6.01494$\\ \hline
  $\displaystyle R_2(m)/S_1(m)$&$\displaystyle 1.61009$&$\displaystyle 1.90731$&$\displaystyle 1.96234$&$\displaystyle 1.97926$&$\displaystyle 1.98754$&$\displaystyle 1.99115$\\ \hline
  \end{tabular}\label{numericalresult2}
\end{table}
%%%%%%%%%%%%%%%%%%%%%%%%%%%%%%%
\begin{table}[H]
  \centering
  \caption{The numerical analysis of $R_3(m)/R_2(m)$}
  \begin{tabular}{|c||c|c|c|c|c|c|}\hline
  $\displaystyle m/H$&$\displaystyle 5.0$&$\displaystyle 7.0$&$\displaystyle 9.0$&$\displaystyle 11.0$&$\displaystyle 13.0$&$\displaystyle 15.0$\\ \hline\hline
  $\displaystyle R_3(m)/R_2(m)$&$\displaystyle -24.137$&$\displaystyle -24.242$&$\displaystyle -23.3081$&$\displaystyle -22.9589$&$\displaystyle -22.7956$&$\displaystyle -22.7469$\\ \hline
  \end{tabular}\label{numericalresult3}
\end{table}
%%%%%%%%%%%%%%%%%%%%%%%%%%%%%%%%%%%%

%%%%%%%%%%%%%%%%%%%%%%%%%%%%%%%%%%%%%%%%%
\subsection{The large mass behavior 
%$m\to\infty$ limit
}
\label{subsection:largemassbehavior}

\begin{figure}[H]
  \centering
  \includegraphics[width=11cm]{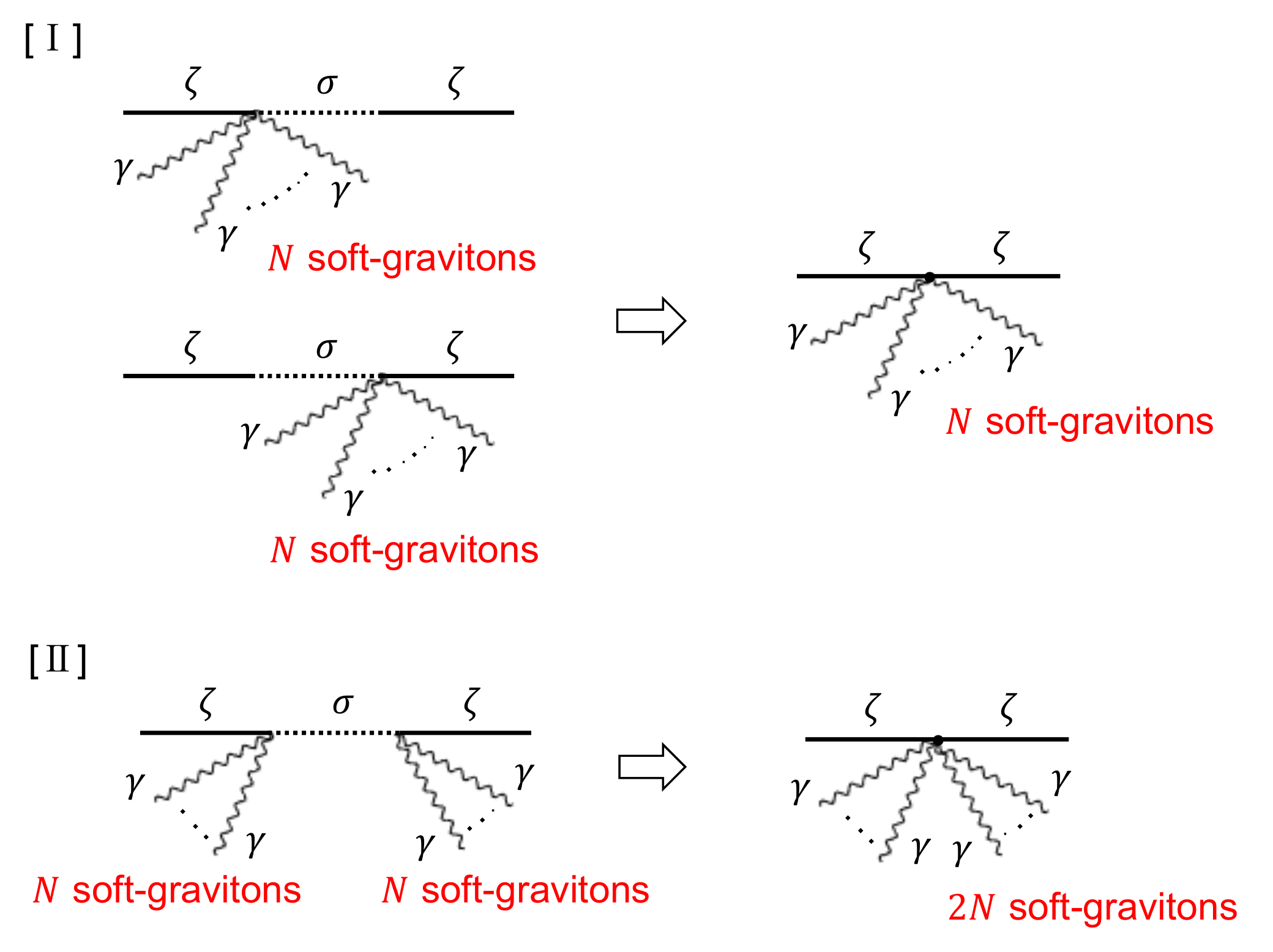}
  \caption{The simplified diagrams when $m\to\infty$}
  \label{simplifiedgene}
\end{figure}

The strategy to study  the asymptotic behaviors of $R_{N}(m)$ and $S_{N}(m)$ for large  $m$ is almost 
 the same  as in Section \ref{sec:largemassbehavior}. 
%  \ref{mtoinftysec}.
We suppose that the dynamical degrees of freedom of $\sigma $  are integrated out 
 in Figure \ref{General} and we introduce effective interaction Hamiltonian as illustrated in 
Figure \ref{simplifiedgene}.
Starting from the interaction Hamiltonian 
(\ref{eq:interactionofNgravitons}), we are able to read off  effective interactions corresponding to 
diagrams on the right hand side in Figure \ref{simplifiedgene}. Namely we get 
\begin{equation}\label{hamihamihami1}
%\textrm{[I]}\ 
H_{I}(t) 
\Bigg \vert _{{\rm Fig.} \ref{simplifiedgene}  \: [{\rm I}]}
=(-1)^{N} 4\frac{C_0 C_{N}}{H^2} a^{3-2N}(t)f(m)\int d^3 x\ \zetadot\gaij\cdots\gamma_{\alpha\beta}\partial_{i}\partial_{j}\cdots\partial_{\alpha}\partial_{\beta}\zetadot\ \times2,
\end{equation}
\begin{equation}
\begin{split}
%\textrm{[I\hspace{-.1em}I]}\ 
H_{I}(t)
\Bigg \vert _{{\rm Fig.} \ref{simplifiedgene}  \: [{\rm II}]}
=4\frac{C_{N}^2}{H^2}a^{3-4N}(t)f(m)\int d^3 x\ \gaij&\cdots\gamma_{\alpha\beta}\partial_{i}\partial_{j}\cdots\partial_{\alpha}\partial_{\beta}\zetadot\\
&\times\gamma_{i'j'}\cdots\gamma_{\alpha'\beta'}\partial_{i'}\partial_{j'}\cdots\partial_{\alpha'}\partial_{\beta'}\zetadot.
\end{split}\label{hamihamihami2}
\end{equation}\vspace{2mm}
for [I] and [II] in Figure \ref{simplifiedgene}, respectively. 
Note that the  function $f(m)$, whose explicit form is not specified here,  is  common 
both  in (\ref{hamihamihami1}) and 
(\ref{hamihamihami2}). 
%%%%%%%%%%%%%%%%%%%%%%%%%%5
%%%%%%%%%%%%%%%%%%%%%%%%%%%

Once we have the effective interactions  (\ref{hamihamihami1}) and 
(\ref{hamihamihami2}), it is almost straightforward to work out  the correlation functions, which turn out to be   
%\leftline{
%\underline{\textbf{[I] Computation of the `$(N,0)$ soft-gravitons' diagram {\sl (simplified)}}}
%}
\begin{equation}
\begin{split}\nonumber
\langle\gamma^{s_1}(\vq_1)\cdots&\gamma^{s_{N}}(\vq_{N})\zeta(\vk_1)\zeta(\vk_2)\rangle
\bigg \vert _{{\rm Fig. } \: \ref{simplifiedgene}\: [ {\rm I} ] }
\\
&=(2\pi)^3\delta^3 (\vq_1+\cdots +\vq_{N}+\vk_1+\vk_2)\\
&\qquad\times\epsilon_{ij}^{s_1}(\vq_1)\cdots\epsilon_{\alpha\beta}^{s_{N}}(\vq_{N})(-i)^2\cdots(-i)^2 (k_2)_{i}(k_2)_{j}\cdots(k_2)_{\alpha}(k_2)_{\beta}\\
&\qquad\times f(m)(-1)^{N}4\frac{C_0 C_{N}}{H^2}\frac{H^2}{q_1^3}\cdots\frac{H^2}{q_{N}^3}\left(\biarii\right)\left(\biari\right)\times2\\
&\qquad\quad
\times\,2!\times N!
\\
&\qquad\quad\times2\,\textrm{Im}\left[\int_{-\infty}^{0}\frac{d\eta}{(\eta H)^{4-2N}}\eta^2 e^{ik_1\eta}\eta^2 e^{ik_2\eta}\right]
\end{split}
\end{equation}\vspace{2mm}
\begin{equation}
\begin{split}
&=(2\pi)^3\delta^3 (\vq_1+\cdots +\vq_{N}+\vk_1+\vk_2)\\
&\qquad\times\epsilon_{ij}^{s_1}(\vq_1)\cdots\epsilon_{\alpha\beta}^{s_{N}}(\vq_{N})\frac{(k_2)_{i}(k_2)_{j}}{(k_2)^2}\cdots\frac{(k_2)_{\alpha}(k_2)_{\beta}}{(k_2)^2}\\
&\qquad\times P_{\gamma}(q_1)\cdots P_{\gamma}(q_{N})\ P_{\zeta}(k_2)\frac{C_0 C_{N}}{\epsilon H^{4-2N}}\,\tilde{R}_{N}(m),
\end{split}
\label{eq:515}
\end{equation}

%\leftline{\underline{\textbf{[I\hspace{-.1em}I] Computation of the `$(N,N)$ soft-gravitons' diagram {\sl
%(simplified)}}}}
\begin{equation}
\begin{split}\nonumber
\langle\gamma^{s_1}(\vq_1)\cdots&\gamma^{s_{2N}}(\vq_{2N})\zeta(\vk_1)\zeta(\vk_2)\rangle
\bigg \vert _{{\rm Fig. } \: \ref{simplifiedgene}\: [ {\rm II } ] }
\\
&=(2\pi)^3\delta^3 (\vq_1+\cdots +\vq_{2N}+\vk_1+\vk_2)\\
&\qquad\times\epsilon_{ij}^{s_1}(\vq_1)\cdots\epsilon_{\alpha\beta}^{s_{2N}}(\vq_{2N})(-i)^2\cdots(-i)^2 (k_2)_{i}(k_2)_{j}\cdots(k_2)_{\alpha}(k_2)_{\beta}\\
&\qquad\times f(m)\,4\frac{C_{N}^2}{H^2}\frac{H^2}{q_1^3}\cdots\frac{H^2}{q_{N}^3}\left(\biarii\right)\left(\biari\right)\,
\times\,2!\times(2N)!
\\
&\qquad\times2\,\textrm{Im}\left[\int_{-\infty}^{0}\frac{d\eta}{(\eta H)^{4(1-N)}}\eta^2 e^{ik_1\eta}\eta^2 e^{ik_2\eta}\right]
\end{split}
\end{equation}\vspace{2mm}
\begin{equation}
\begin{split}
&=(2\pi)^3\delta^3 (\vq_1+\cdots +\vq_{2N}+\vk_1+\vk_2)\\
&\qquad\times\epsilon_{ij}^{s_1}(\vq_1)\cdots\epsilon_{\alpha\beta}^{s_{2N}}(\vq_{2N})\frac{(k_2)_{i}(k_2)_{j}}{(k_2)^2}\cdots\frac{(k_2)_{\alpha}(k_2)_{\beta}}{(k_2)^2}\\
&\qquad\times P_{\gamma}(q_1)\cdots P_{\gamma}(q_{2N})\ P_{\zeta}(k_2)\frac{C_{N}^2}{\epsilon H^{4(1-N)}}\,\tilde{S}_{N}(m).
\end{split}
\label{eq:516}
\end{equation}
Here we have introduced two notations
\begin{eqnarray}
\label{Itildef}
& & 
\tilde{R}_{N}(m)\equiv H^2 f(m)\times(-1)^{N+1}\frac{N!(2N)!}{2^{2N-2}}\: , 
\\
& & 
\label{Jtildef}
\tilde{S}_{N}(m)\equiv H^2 f(m)\times(-1)^{2N+1}\frac{(2N)!(4N)!}{2^{4N-1}}\: .
\end{eqnarray}

Comparison of  these results with (\ref{geneoneresres}) and (\ref{genetworesres}) tells us immediately that 
$\tilde{R}_{N}(m)$ and $\tilde{S}_{N}(m)$ correspond respectively to $R_{N}(m)$ and $S_{N}(m)$ in the 
$m\to\infty$ limit. 
Using (\ref{Itildef}) and (\ref{Jtildef}), we can find the relations:
\begin{eqnarray}
\frac{\tilde{R}_{N+1}(m)}{\tilde{R}_{N}(m)}&=&-\frac{(N+1)^2 (2N+1)}{2},\label{relationII}\\
\frac{\tilde{R}_{2N}(m)}{\tilde{S}_{N}(m)}&=&2.\label{relationIJ}
\end{eqnarray}
Note that the relation (\ref{relationIJ}) is trivial because this `2' is produced by the fact that there are two original diagrams in [I] of Figure \ref{simplifiedgene}.
On the other hand, the relation (\ref{relationII}) is important in the sense that  it serves as a consistency relation  
 relating   $\langle\gamma^{s_1}\cdots\gamma^{s_{N+1}}\zeta\zeta\rangle$ to $\langle\gamma^{s_1}\cdots\gamma^{s_{N}}\zeta\zeta\rangle$ in the $m\to\infty$ limit in this model.
This relation could be useful when searching for new particles whose masses are much greater
 than $10^{14}$ GeV.\vspace{2mm}

Finally we would like to check   the numerical analysis done before.
When $N=1$, (\ref{relationII}) and (\ref{relationIJ}) become
\begin{equation}
\frac{\tilde{R}_{2}(m)}{\tilde{R}_{1}(m)}=-6,\quad\frac{\tilde{R}_{2}(m)}{\tilde{S}_{1}(m)}=2,
\label{eq:R2R1R2S1}
\end{equation}
which are consistent with Table \ref{numericalresult2}.
Note that as a consequence of (\ref{eq:R2R1R2S1})  we get 
$\tilde{R}_{1}(m)/ \tilde{S}_{1}(m) = - 1/3$ which  is nothing but the relation 
(\ref{G23rela}).
If we put  $N=2$ in  (\ref{relationII}), it  becomes
\begin{equation}
\frac{\tilde{R}_{3}(m)}{\tilde{R}_{2}(m)}=-\frac{45}{2}.
\end{equation}
This is also consistent with Table \ref{numericalresult3}. 
% although the numerical value is unstable when we set much larger $m$ by Mathematica 11.

%%%%%%%%%%%%%%%%%%%%%%%%%%%%%%%%%%%%%%%%%%%%%%%%%%%
\section{Summary}
\label{sectionsummary}

In the present paper we have searched for  
 the possibility that the cosmological data could be useful to get hold of unknown heavy particles 
whose masses are on the order of the Hubble parameter $H$ during inflation.  To be more specific 
we considered the quasi-single field inflation model \cite{Quasi} 
and 
tried to develop the methodology of getting signatures of the isocurvaton $\sigma$  in the CMB data. 
In contrast with previous similar attempts \cite{Noumi}, we have paid more attention to 
 correlation functions containing tensor perturbation  $\gamma $ than those of scalar perturbation 
$\zeta $ only.  To keep our investigation as general as possible, we made use of the EFT technique 
 for inflation \cite{Effective, Effective2, Effective3}.

The EFT method tells us that there would exist couplings of the type  $\zeta \sigma$ and
 $\gamma \zeta \sigma$ 
and these couplings could produce potentially observable effects on the power spectrum of $\zeta $
and some of the correlation functions.  
We have therefore launched into detailed study of these effects on $\langle \zeta \zeta \rangle$, 
  $\langle \gamma \zeta \zeta \rangle$
and 
 $\langle \gamma \gamma \zeta \zeta \rangle$. Being particularly interested in 
their dependence on the mass of $\sigma $, we have derived concise formulae  
 of these correlation functions in the form amenable to numerical analyses  
as given in (\ref{f1ori}),  (\ref{f2ori}) and (\ref{f3ori}). The $m$-dependences of the correlation functions are 
illustrated 
in Figures  \ref{fonem}, \ref{ftwom} and \ref{fthreem}.
The effects due to $\sigma $ to 
these correlation functions go down to zero quickly as $m \to \infty $, but 
we have noticed that the large $m$ behavior of these effects  has some common nature. 
We gave a simple explanation of the asymptotic behavior by introducing  a   
short-cut method as illustrated  in Figure \ref{effectivefigure}. 
If we could measure the correlation functions 
$\langle \gamma \zeta \zeta \rangle$
and 
$\langle \gamma \gamma \zeta \zeta \rangle$, 
precise enough in future observation, then we could 
determine the strength of couplings 
$\zeta \sigma $ and $\gamma \zeta \sigma $ together with the mass $m$ . 

We  then went one step further to generalize the above consideration to an 
aribitrary number of gravitons.  
Namely by exploiting the EFT method we constructed a coupling of $\zeta $, $\sigma $ and $N$-gravitons, 
and computed  $\langle\gamma^{s_1}\cdots\gamma^{s_{N}}\zeta\zeta\rangle$ and 
$\langle\gamma^{s_1}\cdots \cdots \gamma^{s_{2N}}\zeta\zeta\rangle$
 by using this coupling once and twice, respectively
as shown in Figure \ref{General}.  Formulae of these correlations are summarized in (\ref{geneoneresres}) 
and (\ref{genetworesres}), and their $m$ dependence has been studied numerically. The asymptotic behaviors 
of the correlations for $m \to \infty$ were also studied by a short-cut method as in the $N=1$ case. 
From these analyses we have seen that the peaks of these correlations move to a larger value of $m$ as 
$N$ increases. Therefore multiple-graviton  correlation functions are more appropriate tools to probe 
larger values of the mass $m$.

In the present paper we have taken up a particular model,  i.e., the quasi-single field inflation model in order 
to make our analyses as definite as possible. In this case, therefore, the heavy particle is $\sigma $  and is
necessarily spinless.   From the standpoint of developing a technique 
to probe unknown heavy particles in the cosmological data, however, it is more desirable to be able to 
handle higher-spin particles as generally as possible. At present we do not have much to say about 
higher-spin case, but hopefully our present analyses could offer a clue for such an extension.

%%%%%%%%%%%%%%%%%%%%%%%%%%%%%%%%%%%%%%%%%%%%%%%%%%
\section*{Acknowledgements}
The authors would like to thank Allan L. Alinea, Renpei Okabe and Motoki Funakoshi for 
useful discussions on effective field theory of inflation. 
Special thanks are due to Allan L. Alinea for 
a careful reading of the manuscript and for many penetrating comments on the numerical calculation. 
His  constructive suggestions have been essential in the present work.  
Their thanks should also go to Tetsuya Onogi, Norihiro Iizuka, Tetsuya Akutagawa, Tomoya Hosokawa 
and Yusuke Hosomi for many discussions. 

%%%%%%%%%%%%%%%%%%%%%%%%%%%%%%%%%%%%%%%%%%%%%%%%%%%%%%%
\appendix
\section{The integration formulae (I)  }
\label{app:A}

We now perform the integration of the following type 
\begin{equation}\label{easyint}
\int_{0}^{\infty}dx\ x^{l} e^{ix}\hankelm(x),
\end{equation}
 which appears in (\ref{zzresult}), (\ref{gzzresult}) and (\ref{ggzzresult})。
Note that $l$ is either  $-1/2$ or $3/2$, and 
\begin{equation}
\nu=
i\mu=i\sqrt{\frac{m^2}{H^2}-\frac{9}{4}}\:. 
\label{eq:defofnu}
\end{equation}
%\footnote{Be careful not to confuse $m$ appeared in (\ref{diffint}) with the mass of $\sigma$.}.
From here, we assume that $\mu$ is real, which means $m\geq3H/2$.

%%%%%%%%%%%%%%%%%%
%\leftline{\underline{{\bf Computation of (\ref{easyint})}}}\vspace{2mm}
 %%%%%%%%%%%%%%%%%%%%%

 Firstly, we perform the indefinite integration by employing Mathematica 11:
 \begin{equation}
 \begin{split}
 \int &dx\ x^{l} e^{ix}\hankelm(x)\\
 &=x^{1+l-i\mu}\frac{-2^{i\mu}\csch(\pi\mu)}{(1+l-i\mu)\Gamma(1-i\mu)}\,{}_2F_2(a_1, a_1+\frac{1}{2}+l; 2a_1, a_1+\frac{3}{2}+l; z)\\
 &+x^{1+l+i\mu}\frac{2^{-i\mu}(1+\coth(\pi\mu))}{(1+l+i\mu)\Gamma(1+i\mu)}\,{}_2F_2(a_2, a_2+\frac{1}{2}+l; 2a_2, a_2+\frac{3}{2}+l; z),
 \end{split}\label{indef1}
 \end{equation}
where $a_1\equiv1/2-i\mu$, $a_2\equiv1/2+i\mu$ and $z\equiv2ix$. Note that ${}_{p}F_{q}$ is 
the  generalized hypergeometric function
 \begin{equation}
 {}_{p}F_{q}(a_1,\cdots,a_{p};b_1,\cdots,b_{q};z)=\sum_{n=0}^{\infty}\frac{(a_1)_{n}\cdots(a_{p})_{n}}{(b_1)_{n}\cdots(b_{q})_{n}}\frac{z^{n}}{n!},
 \end{equation}
 in which $(a)_{n}\equiv a(a+1)\cdots(a+n-1)$ for  $n\geq1$ and $(a)_0=1$, and that  (\ref{indef1}) vanishes when $x=0$. 
In Eq.  (\ref{indef1}) we used the notaion  $\csch(\pi\mu)=1/\sinh(\pi\mu)$.
In order to know the large $x$ behavior of (\ref{indef1}) , we have to examine the asymptotic behavior 
of $\pfq$. The leading terms can be obtained by Mathematica 11:
%(the command is `$\textrm{Series}[\pfq(\cdots; z),\{z,\infty,0\}]$')
\begin{equation}
\begin{split}
&\lim_{z\to\infty}\pfq(a, a+\frac{1}{2}+l; 2a, a+\frac{3}{2}+l; z)\\
&=z^{-\frac{1}{2}-a-l}(-1)^{\frac{3}{2}-a-l}2^{-2+2a}\frac{(1+2a+2l)}{\sqrt{\pi}}\frac{\Gamma(\frac{1}{2}+a)\Gamma(-\frac{1}{2}-l)\Gamma(\frac{1}{2}+a+l)}{\Gamma(-\frac{1}{2}+a-l)}\\
&\qquad+e^{z}z^{-1-a}(\cdots)+z^{-a}(\cdots).
\end{split}\label{asymp1}
\end{equation}
Here both of the ellipses $(\cdots)$ are functions of $a$ and $l$. The term $e^{z}z^{-1-a}(\cdots)$ 
can be eliminated using the $i\epsilon$ prescription. In addition, the terms $z^{-a_1}(\cdots)$ and 
$z^{-a_2}(\cdots)$ are canceled by each other in (\ref{indef1}) . 
% (see Appendix \ref{app1}). 
Therefore, we need to consider only the first term in (\ref{asymp1}).  
Substituting it into (\ref{indef1}) and performing some calculations, 
% (see Appendix \ref{app2}), 
we obtain the following result:
\begin{equation}
\int_{0}^{\infty}dx\ x^{l} e^{ix}\hankelm(x)=e^{\frac{\pi\mu}{2}}\frac{(i/2)^{l}}{\sqrt\pi}\frac{\Gamma(1+l-i\mu)\Gamma(1+l+i\mu)}{\Gamma(l+\frac{3}{2})}. 
\label{resulteasy}
\end{equation}
For $l=-1/2$ and $l=3/2$, (\ref{resulteasy}) turns out to be 
%\leftline{\underline{$l=-1/2$}}
\begin{equation}
\int_{0}^{\infty}dx\ x^{-\frac{1}{2}} e^{ix}\hankelm(x)=\frac{\sqrt{\pi}\,e^{\frac{\pi\mu}{2}}}{\cosh(\pi\mu)}(1-i)
\end{equation}
and
%\leftline{\underline{$l=3/2$}}
\begin{equation}
\begin{split}
\int_{0}^{\infty}dx&\ x^{\frac{3}{2}} e^{ix}\hankelm(x)\\
&=\frac{\sqrt{\pi}\,e^{\frac{\pi\mu}{2}}}{\cosh(\pi\mu)}(-1+i)\times\frac{1}{8}\left(\frac{1}{4}+\mu^2\right)\left(\frac{9}{4}+\mu^2\right), 
\end{split}
\label{resulteasy2}
\end{equation}
respectively.

%%%%%%%%%%%%%%%%%%%%%%%%%%%%%%%%%%%%%%%%%%%%
\section{The integration formulae (II) }
\label{app:B}

We now turn to the evaluation of the integral 
\begin{equation}
\label{diffint}
\int_{0}^{\infty}dx\ x^{m} e^{-ix}\hankelm(x)\int_{x}^{\infty}dy\ y^{l} e^{-iy}\hankelmc(y).
\end{equation}
which also appears in  (\ref{zzresult}), (\ref{gzzresult}) and (\ref{ggzzresult}).
Note that $( l, m) $ is either $(-1/2, -1/2) $, $(3/2, -1/2)$,  $(-1/2, 3/2)$ or $(3/2, 3/2)$, and $\nu$
 is given by  (\ref{eq:defofnu}).
\footnote{The reader has to be careful not to confuse $m$ that appears in (\ref{diffint}) with 
the mass of $\sigma$.}
As before  we assume that $\mu$ is real, or equivalently  $m\geq3H/2$.

%%%%%%%%%%%%%%%%%%%%%%%%%%%%%%%%%
%\leftline{\underline{{\bf Computation of (\ref{diffint})}}}\vspace{2mm}
%%%%%%%%%%%%%%%%%%%%%%%%%%%%%%%%%%
We use the `resummation' trick \cite{chenwang}. Firstly, we rewrite (\ref{diffint}) using the definition of the Hankel function as
\begin{equation}
\begin{split}
\int_{0}^{\infty}dx\ &x^{m} e^{-ix}\hankelm(x)\,\mathcal{I}(x)\\
&=\left[1+\coth(\pi\mu)\right]\int_{0}^{\infty}dx\ x^{m}e^{-ix}\bessel(x)\,\mathcal{I}(x)\\
&\qquad-\csch(\pi\mu)\int_{0}^{\infty}dx\ x^{m}e^{-ix}\besselc(x)\,\mathcal{I}(x),
\end{split}\label{hankelmid}
\end{equation}
where we have defined 
\begin{equation}
\mathcal{I}(x)\equiv\int_{x}^{\infty}dy\ y^{l} e^{-iy}\hankelmc(y).
\end{equation}
Then, using the series expansions
\begin{equation}
\bessel(x)=\sum_{n=0}^{\infty}a_{2n}x^{2n}, \qquad
%\textrm{where}
\quad a_{2n}=x^{i\mu}\,2^{-i\mu}\frac{(-1)^{n}\,2^{-2n}}{n!\,\Gamma(1+n+i\mu)},
\end{equation}
\begin{equation}
e^{-ix}=\sum_{n=0}^{\infty}b_{n}x^{n}, \qquad
%\textrm{where}
\quad b_{n}=\frac{(-i)^{n}}{n!},
\end{equation}
we rewrite the integrand  in (\ref{hankelmid}) also in power series 
\begin{eqnarray}
x^{m}e^{-ix}\bessel(x)&=&\sum_{n=0}^{\infty}\left(\sum_{k=0}^{n}a_{2k}b_{n-2k}\right)x^{n+m}\nonumber\\
&=&\sum_{n=0}^{\infty}\frac{(-i)^{n}\,2^{n+i\mu}\Gamma(\frac{1}{2}+n+i\mu)}{\sqrt{\pi}\,n!\,\Gamma(1+n+2i\mu)}x^{n+m+i\mu}\label{suiko1}\\
&=:&\sum_{n=0}^{\infty}\ \cnp\ x^{n+m+i\mu}.\nonumber
\end{eqnarray}
The coefficient $c_{n}^{+}$ is defined by the last equality. 
Note that the summation in the first line above is performed by Mathematica 11. Similarly, we are able to derive 
\begin{eqnarray}
x^{m}e^{-ix}\besselc(x)&=&\sum_{n=0}^{\infty}\frac{(-i)^{n}\,2^{n-i\mu}\Gamma(\frac{1}{2}+n-i\mu)}{\sqrt{\pi}\,n!\,\Gamma(1+n-2i\mu)}x^{n+m-i\mu}\label{suiko2}\\
&=:&\sum_{n=0}^{\infty}\ \cnm\ x^{n+m-i\mu}.\nonumber
\end{eqnarray}
Here again $c_{n}^{-}$ is defined by the last equality. 
Substituting (\ref{suiko1}) and (\ref{suiko2}) into (\ref{hankelmid}), we end up with 
\begin{equation}
\begin{split}
\int_{0}^{\infty}dx\ x^{m} e^{-ix}&\hankelm(x)\,\mathcal{I}(x)\\
&=\left[1+\coth(\pi\mu)\right]\sum_{n=0}^{\infty} \cnp\int_{0}^{\infty}dx\ x^{n+m+i\mu}\ \mathcal{I}(x)\\
&\qquad-\csch(\pi\mu)\sum_{n=0}^{\infty}\cnm\int_{0}^{\infty}dx\ x^{n+m-i\mu}\ \mathcal{I}(x).
\end{split}\label{tochu10}
\end{equation}

Next, we evaluate the definite integral $\mathcal{I}(x)$ by taking the complex conjugate of the results in  (\ref{resulteasy}) and (\ref{indef1}):
\begin{equation}
\begin{split}
\mathcal{I}(x)&\equiv\int_{x}^{\infty}dy\ y^{l} e^{-iy}\hankelmc(y)\\
&=e^{\frac{\pi\mu}{2}}\frac{(-i/2)^{l}}{\sqrt\pi}\frac{\Gamma(1+l+i\mu)\Gamma(1+l-i\mu)}{\Gamma(l+\frac{3}{2})}\\
&\quad+x^{1+l+i\mu}\frac{2^{-i\mu}\,\csch(\pi\mu)}{(1+l+i\mu)\Gamma(1+i\mu)}\,{}_2F_2(a_2, a_2+\frac{1}{2}+l; 2a_2, a_2+\frac{3}{2}+l; -2ix)\\
&\quad-x^{1+l-i\mu}\frac{2^{i\mu}(1+\coth(\pi\mu))}{(1+l-i\mu)\Gamma(1-i\mu)}\,{}_2F_2(a_1, a_1+\frac{1}{2}+l; 2a_1, a_1+\frac{3}{2}+l; -2ix).
\end{split}\label{eq:tochu0}
\end{equation}
Using the expression (\ref{eq:tochu0}), the integration over $x$ gives us formally the following 
\begin{equation}
\begin{split}
\int_{0}^{\infty}&dx\ x^{n+m+i\mu}\ \mathcal{I}(x)\\
&=e^{\frac{\pi\mu}{2}}\frac{(-i/2)^{l}}{\sqrt\pi}\frac{\Gamma(1+l+i\mu)\Gamma(1+l-i\mu)}{\Gamma(l+\frac{3}{2})}\left.\frac{x^{1+m+n+i\mu}}{1+m+n+i\mu}\right|_{x\to\infty}\\
&\quad+\frac{2^{-i\mu}\,\csch(\pi\mu)}{(1+l+i\mu)\Gamma(1+i\mu)}\int_{0}^{\infty}dx\ x^{1+l+m+n+2i\mu}\pfq(a_2,\cdots; -2ix)\\
&\quad-\frac{2^{i\mu}(1+\coth(\pi\mu))}{(1+l-i\mu)\Gamma(1-i\mu)}\int_{0}^{\infty}dx\ x^{1+l+m+n}\pfq(a_1,\cdots; -2ix).
\end{split}\label{tochu}
\end{equation}
The first term on the right hand side of (\ref{tochu}) is divergent, but this term is actually cancelled by the 
remaining terms.
In order to confirm this, we have to  evaluate the integrals in (\ref{tochu}).

The indefinite integral is known to be 
\begin{equation}
\begin{split}
\int dx\ &x^{p+n}\pfq(a, a+\frac{1}{2}+l; 2a, a+\frac{3}{2}+l; -2ix)\\
&=\frac{(1+2a+2l)\,x^{1+p+n}\,\pfq(a, 1+p+n; 2a, 2+p+n; -2ix)}{(1+p+n)(-1+2a-2p+2l-2n)}\\
&\quad+\frac{2x^{1+p+n}\,\pfq(a, a+\frac{1}{2}+l; 2a, a+\frac{3}{2}+l; -2ix)}{1-2a+2p-2l+2n},
\end{split}\label{tochu2}
\end{equation}
where $p$ is set $p\equiv1+l+m+2i\mu$ for the second term in (\ref{tochu}) 
while $p\equiv1+l+m$ for the third term in (\ref{tochu}).  Note that when $x=0$, the right hand side of (\ref{tochu2}) vanishes.
When $x\to\infty$ on the other hand, we have to know  the asymptotic behavior of the hypergeometric 
functions as before. We can use (\ref{asymp1}) for the second term in (\ref{tochu2}), and after some 
calulations, 
% (see Appendix \ref{app3}), 
we are able to confirm that the second term in (\ref{tochu2}) eliminates the divergent first term 
in (\ref{tochu}).  As for the first term in (\ref{tochu2}), the asymptotic behavior is given by 
\begin{equation}
\begin{split}
&\lim_{z\to\infty}\pfq(a, 1+p+n; 2a, 2+p+n; z)\\
&=z^{-1-p-n}(-1)^{1-p-n}2^{-1+2a}\frac{(1+p+n)}{\sqrt{\pi}}\frac{\Gamma(\frac{1}{2}+a)\Gamma(-1+a-p-n)\Gamma(1+p+n)}{\Gamma(-1+2a-p-n)}\\
&\qquad+e^{z}z^{-1-a}(\cdots)+z^{-a}(\cdots),
\end{split}\label{asymp2}
\end{equation}
and we are allowed to  neglect the second and the third terms of (\ref{asymp2}) for the same reason as (\ref{asymp1}).

When substituting (\ref{asymp2}) into (\ref{tochu2}), and then substituting (\ref{tochu2}) into (\ref{tochu}), the second term in (\ref{tochu}) vanishes. This is because the denominator of (\ref{asymp2}) becomes
\begin{equation}
\Gamma(-1+2a_2-p-n)=\Gamma(-(1+l+m+n)),
\end{equation}
which is divergent since $1+l+m+n$ is an integer greater than or equal to zero (recall that $l$ and $m$ are $-1/2$ or $3/2$). 
Finally, only the third term in (\ref{tochu}) remains, and just by using (\ref{asymp2}), we obtain
\begin{equation}
\begin{split}
\cnp&\int_{0}^{\infty}dx\ x^{n+m+i\mu}\ \mathcal{I}(x)\\
&=\frac{(-1)^{-2(l+m)}(2i)^{-2-l-m}}{\pi\,\sinh(\pi\mu)}\frac{e^{\pi\mu}}{1+m+n+i\mu}\\
&\qquad\quad\times(-1)^{n}\frac{\Gamma(2+l+m+n)}{\Gamma(n+1)}\frac{\Gamma(\frac{1}{2}+n+i\mu)\Gamma(-\frac{3}{2}-l-m-n-i\mu)}{\Gamma(1+n+2i\mu)\Gamma(-1-l-m-n-2i\mu)}.
\end{split}\label{tochup}
\end{equation}
Similarly, we are able to show the formula 
\begin{equation}
\begin{split}
\cnm&\int_{0}^{\infty}dx\ x^{n+m-i\mu}\ \mathcal{I}(x)\\
&=\frac{(-1)^{-2(l+m)}(2i)^{-2-l-m}}{\pi\,\sinh(\pi\mu)}\frac{-1}{1+m+n-i\mu}\\
&\qquad\quad\times(-1)^{n}\frac{\Gamma(2+l+m+n)}{\Gamma(n+1)}\frac{\Gamma(\frac{1}{2}+n-i\mu)\Gamma(-\frac{3}{2}-l-m-n+i\mu)}{\Gamma(1+n-2i\mu)\Gamma(-1-l-m-n+2i\mu)}.
\end{split}\label{tochum}
\end{equation}

Using (\ref{tochup}) and (\ref{tochum}) into (\ref{tochu10}),  we get the following  result:
\begin{equation}
\begin{split}
&
\displaystyle\int_{0}^{\infty}dx\ x^{m} e^{-ix}\hankelm(x)\int_{x}^{\infty}dy\ y^{l} e^{-iy}\hankelmc(y)
  \\
&  \displaystyle\qquad=\frac{(-1)^{-2(l+m)}(2i)^{-2-l-m}}{\pi\,\sinh^2(\pi\mu)}
  \\
&  \displaystyle\qquad\qquad\times\sum_{n=0}^{\infty}(-1)^{n}\frac{\Gamma(2+l+m+n)}{\Gamma(n+1)}
  \\
&
  \displaystyle\qquad\qquad\qquad\times\Biggl[\frac{e^{2\pi\mu}}{1+m+n+i\mu}\frac{\Gamma(\frac{1}{2}+n+i\mu)\Gamma(-\frac{3}{2}-l-m-n-i\mu)}{\Gamma(1+n+2i\mu)\Gamma(-1-l-m-n-2i\mu)}
  \\
&
  \displaystyle\qquad\qquad\qquad\quad+\frac{1}{1+m+n-i\mu}\frac{\Gamma(\frac{1}{2}+n-i\mu)\Gamma(-\frac{3}{2}-l-m-n+i\mu)}{\Gamma(1+n-2i\mu)\Gamma(-1-l-m-n+2i\mu)}\Biggl]
\end{split}
\label{resultdiff}
\end{equation}
Note that in the course of deriving this formula,  we assumed  that  $l+m$ is an integer greater than 
or equal to $-1$.  
By putting $(l, m)=(-1/2, -1/2)$, $(3/2, -1/2))$, $(-1/2, 3/2)$ and $(3/2, 3/2)$ in (\ref{resultdiff}), 
we arrive at the following set of formulae: 
\vskip3mm

\leftline{\underline{$(l, m)=(-1/2,-1/2)$}}
\begin{equation}
\begin{split}
\int_{0}^{\infty}&dx\ x^{-\frac{1}{2}} e^{-ix}\hankelm(x)\int_{x}^{\infty}dy\ y^{-\frac{1}{2}} e^{-iy}\hankelmc(y)\\
&=\frac{1}{\pi\,\sinh(\pi\mu)}\sum_{n=0}^{\infty}(-1)^{n}\Biggl\{\frac{e^{2\pi\mu}}{(\frac{1}{2}+n+i\mu)^2}-\frac{1}{(\frac{1}{2}+n-i\mu)^2}\Biggl\}, 
\end{split}
\label{eq:-1/2-1/2}
\end{equation}

\leftline{\underline{$(l, m)=(3/2, -1/2)$}}
\begin{equation}
\begin{split}
\int_{0}^{\infty}&dx\ x^{-\frac{1}{2}} e^{-ix}\hankelm(x)\int_{x}^{\infty}dy\ y^{\frac{3}{2}} e^{-iy}\hankelmc(y)\\
&=-\frac{1}{4\pi\,\sinh(\pi\mu)}\,\sum_{n=0}^{\infty}(-1)^{n}(n+1)(n+2)\\
&\qquad\qquad\qquad\qquad\times\Biggl\{e^{2\pi\mu}\frac{(1+n+2i\mu)(2+n+2i\mu)}{(\frac{1}{2}+n+i\mu)^2(\frac{3}{2}+n+i\mu)(\frac{5}{2}+n+i\mu)}\\
&\qquad\qquad\qquad\qquad\qquad\qquad-\frac{(1+n-2i\mu)(2+n-2i\mu)}{(\frac{1}{2}+n-i\mu)^2(\frac{3}{2}+n-i\mu)(\frac{5}{2}+n-i\mu)}\Biggl\}, 
\end{split}
\label{eq:3/2-1/2}
\end{equation}

\leftline{\underline{$(l, m)=(-1/2, \,3/2)$}}
\begin{equation}
\begin{split}
\int_{0}^{\infty}&dx\ x^{\frac{3}{2}} e^{-ix}\hankelm(x)\int_{x}^{\infty}dy\ y^{-\frac{1}{2}} e^{-iy}\hankelmc(y)\\
&=-\frac{1}{4\pi\,\sinh(\pi\mu)}\,\sum_{n=0}^{\infty}(-1)^{n}(n+1)(n+2)\\
&\qquad\qquad\qquad\qquad\times\Biggl\{e^{2\pi\mu}\frac{(1+n+2i\mu)(2+n+2i\mu)}{(\frac{1}{2}+n+i\mu)(\frac{3}{2}+n+i\mu)(\frac{5}{2}+n+i\mu)^2}\\
&\qquad\qquad\qquad\qquad\qquad\qquad-\frac{(1+n-2i\mu)(2+n-2i\mu)}{(\frac{1}{2}+n-i\mu)(\frac{3}{2}+n-i\mu)(\frac{5}{2}+n-i\mu)^2}\Biggl\}, 
\end{split}
\label{eq:-1/23/2}
\end{equation}

\leftline{\underline{$(l, m)=(3/2, \,3/2)$}}
\begin{equation}
\begin{split}
\int_{0}^{\infty}&dx\ x^{\frac{3}{2}} e^{-ix}\hankelm(x)\int_{x}^{\infty}dy\ y^{\frac{3}{2}} e^{-iy}\hankelmc(y)\\
&=\frac{1}{16\pi\,\sinh(\pi\mu)}\\
&\quad\quad\,\times\sum_{n=0}^{\infty}(-1)^{n}(n+1)(n+2)(n+3)(n+4)\\
&\qquad\times\Biggl\{e^{2\pi\mu}\frac{(1+n+2i\mu)(2+n+2i\mu)(3+n+2i\mu)(4+n+2i\mu)}{(\frac{1}{2}+n+i\mu)(\frac{3}{2}+n+i\mu)(\frac{5}{2}+n+i\mu)^2(\frac{7}{2}+n+i\mu)(\frac{9}{2}+n+i\mu)}\\
&\qquad\qquad-\frac{(1+n-2i\mu)(2+n-2i\mu)(3+n-2i\mu)(4+n-2i\mu)}{(\frac{1}{2}+n-i\mu)(\frac{3}{2}+n-i\mu)(\frac{5}{2}+n-i\mu)^2(\frac{7}{2}+n-i\mu)(\frac{9}{2}+n-i\mu)}\Biggl\}. 
\end{split}
\label{eq:3/23/2}
\end{equation}
%%%%%%%%%%%%%%%%%%%%%%%%%%%%%
%{\color{red}{
\section{The integration formulae (III)} 
\label{app:C}

In the course of deriving Eqs. (\ref{f1til}), (\ref{f2til}) and (\ref{f3til}), we encounter 
 the integration of the following type
\begin{eqnarray}
\int _{-\infty} ^{0} d\eta \: \eta ^{2N} e^{2ik \eta}
=
-\left ( \frac{-1}{2k}\right )^{1+2N}
 \int _{0}^{+\infty} dx \: x^{2N}e^{-ix}.
\end{eqnarray}
Here we have changed the integration variable from $\eta$ to  $x=-2k\eta $. For large $x$, the integrand is 
oscillating and we employ the ``$i \varepsilon $ prescription" to render the integration well-defined, i.e., 
\begin{eqnarray}
\int _{0}^{+\infty} dx \: x^{2N} e^{-ix} 
\Rightarrow 
\lim_{\varepsilon \to +0}\int _{0}^{+\infty} dx\: x^{2N} e^{-i(1-i\varepsilon)x}
%=\lim_{\varepsilon \to +0}\int _{0}^{+\infty} dx\: x^{2N} e^{-ix} e^{-\varepsilon x} .
\end{eqnarray}
%%%%%%%%%%%%%%%%%%%%%%%%%%%
%%%%%%%%%%%%%%%%%%%%%%%%%
The $x$-integration is easily evaluated by the following formula
\begin{eqnarray}
\int _{0}^{+\infty} dx\: x^{2N} e^{-i(1-i\varepsilon)x}
&=&
\left ( i \frac{d}{dt}\right )^{2N} \int _{0}^{ + \infty} dx\: e^{-itx} \Bigg \vert _{t=1-i\varepsilon} 
\nonumber \\
&=&
\left ( i \frac{d}{dt} \right )^{2N} \frac{1}{it} \Bigg \vert _{t=1-i\varepsilon} 
\nonumber \\
&=&
i (-1)^{N+1} \frac{(2N)!}{(1-i \varepsilon )^{2N+1}}
\end{eqnarray}
In this way we arrive at the integration formula
\begin{eqnarray}
\int _{-\infty} ^{0} d\eta \: \eta ^{2N} e^{2ik \eta}
&=&
i(-1)^{N+1}\left ( \frac{1}{2k} \right )^{2N+1}(2N)! . 
\end{eqnarray}
This formula is also useful to derive Eq. (\ref{eq:515}) and (\ref{eq:516}).
%}}
%%%%%%%%%%%%%%%%%%%%%%%%%%%%%%%
%\newpage

\end{document}